\documentclass{article}

\usepackage{iclr2026_conference,times}
\iclrfinalcopy

\usepackage{comment}

\usepackage[utf8]{inputenc} 
\usepackage[T1]{fontenc}    
\usepackage{hyperref}       
\usepackage{url}            
\usepackage{booktabs}       
\usepackage{amsfonts}       
\usepackage{multirow}       
\usepackage{nicefrac}       
\usepackage{wrapfig}
\usepackage{microtype}      
\usepackage{xcolor}         
\usepackage{graphicx}
\usepackage{amsmath}
\usepackage{cleveref}
\usepackage{algorithm}
\usepackage{algpseudocode}
\usepackage[T1]{fontenc}      
\usepackage[utf8]{inputenc}   
\usepackage{xcolor}           
\usepackage{listings}
\usepackage{dirtree}
\usepackage{enumitem}
\usepackage{amssymb} 

\usepackage{caption}   

\usepackage{amsmath,amsfonts,bm}

\def\eqref#1{equation~\ref{#1}}

\def\1{\bm{1}}

\DeclareMathAlphabet{\mathsfit}{\encodingdefault}{\sfdefault}{m}{sl}
\SetMathAlphabet{\mathsfit}{bold}{\encodingdefault}{\sfdefault}{bx}{n}

\DeclareMathOperator*{\argmin}{arg\,min}

\definecolor{gruvbox_pink}{RGB}{219, 48, 122}
\definecolor{gruvbox_blue}{RGB}{69, 133, 136}
\definecolor{gruvbox_green}{RGB}{152, 151, 26}
\definecolor{gruvbox_purple}{RGB}{177, 98, 134}
\definecolor{gruvbox_aqua}{RGB}{104, 157, 106}

\definecolor{codegreen}{rgb}{0,0.6,0}
\definecolor{codegray}{rgb}{0.5,0.5,0.5}
\definecolor{codepurple}{rgb}{0.58,0,0.82}
\definecolor{backcolour}{rgb}{0.95,0.95,0.92}

\lstdefinestyle{codestyle}{
    backgroundcolor=\color{backcolour},   
    commentstyle=\color{gruvbox_green},
    keywordstyle=\color{gruvbox_pink},
    numberstyle=\tiny\color{codegray},
    stringstyle=\color{gruvbox_purple},
    basicstyle=\ttfamily\tiny,  
    breakatwhitespace=false,         
    breaklines=true,                 
    captionpos=b,                    
    keepspaces=true,                 
    numbers=left,                    
    numbersep=5pt,                  
    showspaces=false,                
    showstringspaces=false,
    showtabs=false,                  
    tabsize=2,
    frame=single,
    framesep=3pt
}

\usepackage{listings}
\usepackage{xcolor}
\usepackage{tcolorbox}
\tcbuselibrary{listings,skins,breakable}
\usepackage{inconsolata} 

\definecolor{codebg}{RGB}{255,255,255}      
\definecolor{codekeyword}{RGB}{175,0,219}   
\definecolor{codestring}{RGB}{206,145,120}  
\definecolor{codecomment}{RGB}{106,153,85}  
\definecolor{codetext}{RGB}{0,0,0}          
\definecolor{codebuiltin}{RGB}{0,128,255}   
\definecolor{codeself}{RGB}{0,128,255}      
\definecolor{framecolor}{RGB}{150,150,150}  

\lstdefinestyle{modernpython}{
    language=Python,
    basicstyle=\ttfamily\scriptsize\color{codetext},
    numbers=none,
    keywordstyle=\color{codekeyword}\bfseries,
    commentstyle=\color{codecomment},
    stringstyle=\color{codestring},
    showstringspaces=false,
    breaklines=true,
    tabsize=4,
    keepspaces=true,
    columns=flexible,
    morekeywords={super, self, pass, from, import, as, def, class, None, int},
    emph={RMSNorm, BaseMLP, BaseRotaryEmbedding, BaseAttention, BaseDecoderLayer, 
          LlamaConfig, rotate_half, apply_rotary_pos_emb, repeat_kv, 
          eager_attention_forward, LlamaRMSNorm, LlamaRotaryEmbedding, 
          LlamaMLP, LlamaAttention, shared_library},
    emphstyle=\color{codebuiltin},
}

\newtcblisting{moderncode}{
    listing only,
    listing engine=listings,
    listing options={style=modernpython},
    enhanced,
    arc=4mm,              
    boxrule=1pt,        
    colframe=framecolor,  
    colback=codebg,       
    left=8pt,
    right=8pt,
    top=2pt,
    bottom=2pt,
}

\tcbset{
    modernstyle/.style={
        listing only,
        listing engine=listings,
        listing options={style=modernpython},
        enhanced,
        rounded corners,      
        boxrule=0.75pt,
        colframe=framecolor,
        colback=codebg,
        left=8pt,
        right=8pt,
        top=8pt,
        bottom=8pt,
    }
}

\newcommand{\program}{\ensuremath{\rho}}
\newcommand{\library}{\ensuremath{\mathcal{L}}}
\newcommand{\loss}[1]{\ensuremath{\ell}\left( #1 \right)}
\newcommand{\sample}[1]{\ensuremath{\textsc{Sample}\left( #1 \right)}}
\newcommand{\samplek}[1]{\ensuremath{\textsc{Sample}_K\left( #1 \right)}}

\newcommand{\NAME}{\textsc{Librarian}}
\newcommand{\BENCH}{\textsc{MiniCode}}
\newcommand{\ITEM}{file}

\title{Refactoring Codebases through\\ Library Design }

\author{%
  Žiga Kovačič\\
  Cornell University\\
  \texttt{zk66@cornell.edu}
  \And
  Justin T Chiu\\
  Cohere\\
  \texttt{chiu.justin.t@gmail.com} 
  \And
  Celine Lee\\
  Cornell University\\
  \texttt{cl923@cornell.edu}
  \And
  Wenting Zhao\\
  Cornell University\\
  \texttt{wz346@cornell.edu}
  \And
  Kevin Ellis\\
  Cornell University\\
  \texttt{kellis@cornell.edu}
  \\
}

\begin{document}

\maketitle

\begin{abstract}

 Maintainable and general software allows developers to build robust applications efficiently, yet achieving these qualities often requires refactoring specialized solutions into reusable components. This challenge becomes particularly relevant as code agents become used to solve isolated one-off programming problems. We investigate code agents' capacity to refactor code in ways that support growth and reusability. We first investigate what makes a good refactoring, finding via simulation results and a human study that Minimum Description Length best correlates with preferable refactorings. We then present both a benchmark and a method for refactoring: \textsc{MiniCode}, a benchmark where multiple files must be refactored into a shared library, and \textsc{Librarian}, a sample-and-rerank method for generating reusable libraries. We compare \textsc{Librarian} to state-of-the-art library generation methods, and study it on real-world code bases.

\end{abstract}

\section{Introduction}

Writing code is mainly a matter of \emph{re}writing code: debugging, refactoring, optimizing, and other activities within the software engineering lifecycle.
But poor rewrites incur technical debt, with such debt costing up to \$2 \emph{trillion} annually~\citep{Tews}.
This problem will likely worsen as language models become increasingly responsible for generating code, because they excel at solving isolated programming problems, but their context length demands a myopic view of the codebase.
Effective code refactoring at scale is a design problem whose concerns center on re-usability and maintainability. A classic example illustrates this design challenge: Human programmers often create overly-specialized, redundant solutions to similar problems and would benefit from redesigning specialized solutions into a shared library. 

\begin{figure}
\centering
\includegraphics[width=\textwidth]{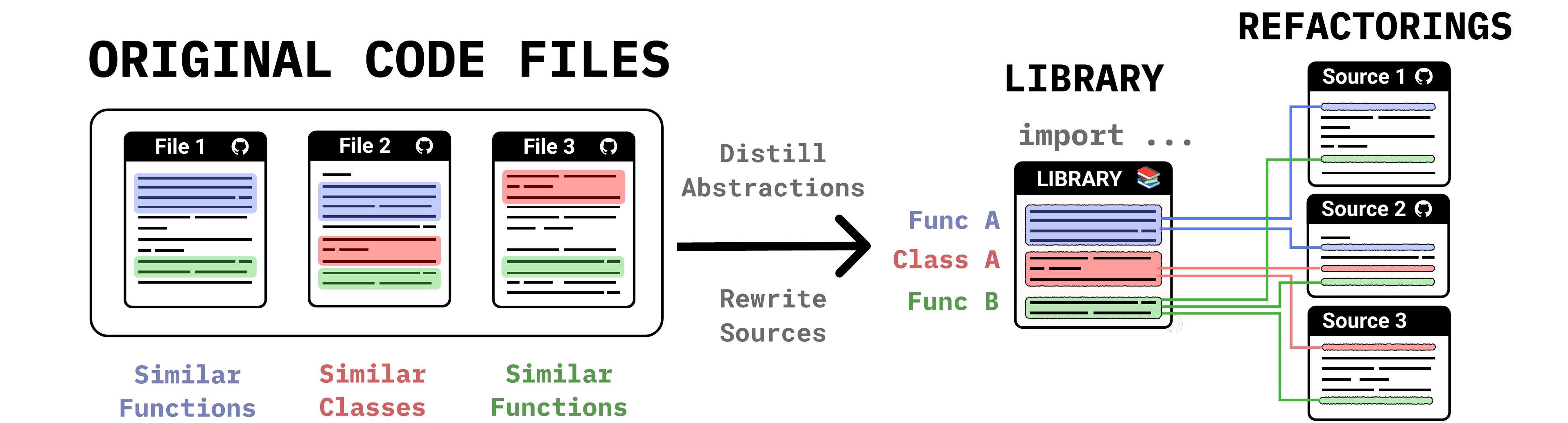}
\vspace{-10pt}
\caption{Overview of the refactoring problem.
A refactoring task comprises a set of  \ITEM{}s.
We refactor the \ITEM{}s by designing a new library.
Candidate refactorings are evaluated based on a refactoring metric, and are expected to maintain correctness of the original code sources (pass rate).
We explore several refactoring metrics in this paper.
}\label{fig:overview}
\vspace{-5pt}
\end{figure}

Here we focus on refactoring multiple files into a reusable software library, which raises two questions: (1) How should we quantify the quality of a library refactoring, and (2)
How can we use LLMs to refactor specialized code into reusable libraries?
To answer those question, we develop a new method and a benchmark.
This goes beyond past work in \emph{library learning}~\citep{Wong2021LeveragingLT,10.5555/3692070.3693967,10.1145/3453483.3454080,10.1145/3571234,Dechter:2013:BLV:2540128.2540316,Grand2023LILOLI,liang2010learning} which synthesizes subroutines across small programs in i.e. $\lambda$-calculus, because instead we address the more naturalistic problem of redesigning code written in contemporary high-level languages, such as Python, producing classes, methods, and helper functions in the style of a human-written library.
We develop a method, \NAME{} (\Cref{fig:overview}), which samples possible code rewrites and then reranks those samples based on criteria designed to capture a good refactoring.
To generate potential rewrites, we develop methods for clustering pieces of code together that share common structure so that a succinct prompt can rewrite them jointly into their refactored form.
To find strong criteria for ranking potential rewrites, we study a variety of metrics across machine learning and software engineering, both on programming benchmarks and via a human study.

\newpage

To evaluate our method and systematically assess the capability of current agents to generate libraries, we introduce a new benchmark, \BENCH, which addresses three key desiderata missing from existing benchmarks. First, open-ended design: unlike SWE-Bench \citep{jimenez2024swebench}, Commit0 \citep{zhao2025commit}, and RefactorBench \citep{gautam2025refactorbench} which primarily focus on functional correctness, \BENCH{} presents an unconstrained library design problem. Agents create a library that can be imported back into a repository, with complete freedom to design the interface and implementation from scratch—optimizing for software engineering objectives like reusability and maintainability. Second, verifiability: we ensure objective evaluation by retaining the unit tests from all repositories that will import the designed library, allowing us to verify that the solutions work correctly across multiple use cases. Third, large context: agents must understand and synthesize information from multiple code sources (files) simultaneously to design a unified library that consolidates specialized code sources into a general interface. 

We contribute the following:
\begin{enumerate}[wide, labelwidth=!, labelindent=0pt]
    \item Study of different metrics for library creation via a user study and program synthesis benchmarks, with the surprising finding that neural language models offer a stronger signal of refactoring quality than classic metrics from the software engineering community.
    \item A new benchmark, \BENCH{}, covering both competition programming and real-world machine learning programs from the Transformers and Diffusers libraries.
    \item A new algorithm, \NAME{}, which outperforms prior library learners on existing benchmarks and scales to complex codebases.
    \item Refactorings of real-world codebases widely used in the machine learning community. Our method successfully refactors the Transformers and Diffusers libraries from Huggingface to be shorter and more reusable.
    To the best of our knowledge this is the first time library learning has have been successfully applied to real-world software projects.
\end{enumerate}


\section{Related work}
\paragraph{Library Learning.} Systems which perform library learning research discover shared abstractions across a large number of small programs, which they use to automatically define new subroutines.
Systems such as DreamCoder~\citep{10.1145/3453483.3454080}, Trove~\citep{wang2024troveinducingverifiableefficient}, LiLo~\citep{Grand2023LILOLI}, and REGAL~\citep{10.5555/3692070.3693967} automatically construct such libraries with the goal of making future program synthesis tasks easier to solve, once the learned library is in hand.
Our work is closest to REGAL~\citep{10.5555/3692070.3693967}, which clusters related code and refactors using language models.
However, existing library learning approaches have primarily been demonstrated in small-scale, constrained domains, limiting their applicability to typical software engineering tasks, such as consolidating multiple repositories into cohesive libraries. By framing library learning within the context of realistic software development, we expand the relevance of library learning to the everyday practice of software engineering.

\paragraph{Repo-level coding benchmarks.}
Recent work has explored the application of language models to repository-level software engineering tasks. Existing benchmarks include SWE-bench~\citep{jimenez2024swebench}, which evaluates models on their ability to resolve real-life GitHub issues, and Commit-0 \citep{zhao2025commit}, which requires agents to fill in function definitions. Such benchmarks primarily evaluate functional correctness via unit tests, without assessing the quality or maintainability of the resulting codebase. 
RefactorBench~\citep{gautam2025refactorbench} takes a step in this direction by benchmarking the ability to follow specific refactoring instructions.
Our work differs by requiring models to perform a more open-ended task:
Redesigning code to be more modular and compact by discovering reused abstractions, while retaining verifiability by re-using downstream unit tests. Additionally, libraries must be created without any scaffolding limitations such as preexisting function definitions affording more design freedom than Commit-0.

\paragraph{Program optimization.}
While our goal is to optimize the quality of libraries, other works focus on improving execution speed through correctness-preserving transformations
\citep{waghjale2024ecco,ouyang2025kernelbenchllmswriteefficient, 10.1145/2451116.2451150}.
Both forms of program optimization, compression and speed, are more open-ended than optimizing only for correctness, as there does not exist a ground-truth answer.
Prior work on program optimization benchmarks study code at the file level. We propose a benchmark that transforms programs at a larger scale, across multiple code files.

 \section{Problem Statement} 
\label{sec:setup}
Given $N$ related files $\left\{ \program_n \right\}_{n=1}^N$, the goal is to create a library $\library$ that captures shared abstractions.
The original files $\left\{ \program_n \right\}_{n=1}^N$, which define the refactoring problem, we call a \textbf{task}.
The new library must
support all original use cases in the task by extracting latent shared abstractions. 
This is accomplished by searching for refactorings that are both correct and `natural'. Correctness is straightforward to define via unit tests, but naturalness is more challenging to quantify.

Shorter code is potentially simpler and less redundant. One potential metric is to count the number of tokens, lines-of-code, or syntax tree nodes in the proposed library and refactored code ~\citep{Dechter:2013:BLV:2540128.2540316,polozov2015flashmeta,10.1145/3571234,10.1145/3571207}. 
But minimizing program size has obvious failure modes: code should also be understandable and extensible, which can be in tension with merely finding the shortest program.\footnote{\href{https://wiki.c2.com/?PerlGolf}{Perl Golf} is a game where participants attempt to write the shortest Perl program accomplishing a given task. The resulting code is famously incomprehensible, even by the standards of Perl.}
Other work in program synthesis \citep{DBLP:conf/icml/LiangJK10,solomonoff1964formal,10.1145/3453483.3454080} instead optimizes \emph{Minimum Description Length} (MDL), or negative log probability under a reference distribution.
In the software engineering community, other metrics such as cyclomatic complexity and maintainability index have been defined for similar purposes:
These are more complex metrics that examine the syntax tree, call graph, and other statically-analyzable structures~\citep{1702388}.
What metric should we use? 
We revisit this question in Section~\ref{sec:metric}, where we empirically compare candidate metrics and human preferences before fixing our choice for the rest of the paper.

For now assume a placeholder metric $M$ measuring refactoring quality; we seek to minimize $M$ while preserving correctness.
Given a task comprising \ITEM{}s $\left\{ \program_n \right\}_{n=1}^N$, we output both a new library $\library$, as well as rewritten refactorings of the original \ITEM{}s, $\left\{ \program'_n \right\}_{n=1}^N$.
We define tests passed $\tau(\program_n)$ as the set of unit tests $\program_n$ passes, and consider both  refactoring several \ITEM{}s ($N>1$) and also refactoring a single large \ITEM{}  ($N=1$).
We optimize the following objective, which prefers passing at least the same tests as the original programs \emph{and} minimizing the chosen metric $M$:
\begin{equation}
\loss{\library, \{ \program_n'\}} =
\begin{cases}
M(\library, \{\program_n'\}) & \forall \program_n , \tau(\program_n)  \subseteq \tau(\program_n') \\
\infty & \text{otherwise}
\end{cases}
\end{equation}

\section{\BENCH{}—Library Design and Refactoring Benchmark} 
\BENCH{} 
presents systems with a task comprising a set of \ITEM{}s, then asks them to refactor the \ITEM{}s into a unified library alongside refactorings of the original \ITEM{}s.
There are two key desiderata for benchmark tasks: They should have related \ITEM{}s sharing latent abstractions, and should also be verifiable, to measure how well refactored \ITEM{}s preserve functional correctness.
We source a variety of problems (\Cref{tab:dataset_characteristics}).


\begin{table}[H]
\centering
\caption{\BENCH{} Statistics}
\begin{tabular}{lccccc}
\toprule
\textbf{Domain} & \textbf{Files} & \textbf{Tasks} & \textbf{Avg LoC} & \textbf{Avg Tests / \ITEM} \\
\midrule
Code Contests \citep{CodeContests}      & 300 & 10 & 87 & 10  \\
Transformers \citep{wolf-etal-2020-transformers}  & 10 & 1 & 538 & 181  \\
Diffusers \citep{von-platen-etal-2022-diffusers} & 11 & 2 & 685 & 75   \\
Logo \citep{Wong2021LeveragingLT} & 300 & 1 & 10 & 1  \\
Date \citep{srivastava2023beyond} & 246 & 1 & 14 & 1  \\
\bottomrule
\end{tabular}
\label{tab:dataset_characteristics}
\end{table}

\paragraph{CodeContests.}
Competition problems are crafted with specific variations of algorithmic approaches in mind, resulting in both shared latent concepts and the required test cases. As a result, competition coding is both verifiable, and ready to refactor.
We therefore take solutions, prompts, and tests from \textsc{CodeContests} \citep{li2022competition}, a competition programming dataset.

\paragraph{Huggingface Transformers Library.} We test refactoring across implementations of large language and vision–language models  from the Huggingface \texttt{transformers} repository (\texttt{modelling\_<name>.py} files, e.g., Qwen2, LLaMA, DeepSeek-V3). Unlike competition coding, these sources are production-scale and Huggingface requires that all changes pass an extensive suite of integration tests before merging into the main branch. A refactoring is only deemed correct if it passes the unmodified Transformers test suite, making this a high-stakes setting that requires correctness and compatibility. 

\paragraph{Huggingface Diffusers Library.} We test refactoring across implementations of diffusion models from the Huggingface \texttt{diffusers} repository (\texttt{unet\_<name>.py} and \texttt{scheduler\_<name>.py} files, e.g., Stable Diffusion UNet, DDPMScheduler), yielding two distinct tasks. Like Transformers, Diffusers requires that all changes pass a comprehensive suite of integration tests before merging into the main branch.

\paragraph{Logo \& Date.}
The library learning literature already has existing benchmarks:
Typically they seek to learn a single library from a task comprising many sources, and then test that library on holdout program synthesis tasks.
Logo and Date were used in the recent related work REGAL~\citep{10.5555/3692070.3693967}, which we incorporate wholesale to understand how our new method compares to state-of-the-art library learning.
The associated programming problems were created by humans, but their solutions were generated by gpt-3.5-turbo.

\section{\NAME: Refactoring Code to Create Libraries}
\label{sec:method}
\NAME{} generates a new library from a set of \ITEM{}s, while migrating the \ITEM{}s to use that new library (Figure~\ref{fig:overview}), following a sample-and-rerank framework:
Prompting a backend LLM or agent to sample $K$ candidates, and picking the one minimizing the loss $\ell$.
Naively,
\begin{align}
\library, \left\{ \program'_n \right\} = \argmin_{\library, \left\{ \program'_n \right\}\in \samplek{\left\{ \program_n \right\}}} 
\loss{\library, \left\{ \program'_n \right\}}
\end{align}
for metric $M$ and sampling budget $K$.
But this cannot work for large tasks with many programs, which would not fit into the context of most LLMs.
Even long context models cannot process the entirety of e.g the Linux kernel, and even if they could, it is not clear that such a strategy is the most efficient way of focusing the language model's attention.
To address this, we wrap sample-and-rerank with a clustering algorithm that decomposes the task into manageable chunks, described next.

\paragraph{Clustering.}

Meaningful abstractions arise when programs share underlying functionality or structure. To surface these, we cluster the task's \ITEM{}s into small groups that are likely to share reusable structure, 
and refactor each cluster separately from the rest.
This decomposition shrinks the prompt size, and gives independent searches for the best per-cluster refactoring, which may be more tractable.

\NAME{}'s clustering extends REGAL~\citep{10.5555/3692070.3693967}, which clusters programs 
by assuming each program is paired with a natural language description of the problem it solves, and clustering embeddings of those descriptions.
Since similar problems need not imply similar solution code, we instead prompt a model to summarize each \ITEM{} and cluster by these summaries. 
Specifically, we define $\textsc{Cluster}_S\left( \left\{ \program_n \right\} \right)$ as performing agglomerative clustering~\citep{ward1963hierarchical} on the task's \ITEM{}s $\left\{ \program_n \right\}$ to produce a set of set of \ITEM{}s, each of which is a cluster of size $S$.
We use text-embedding-ada-002
 to embed descriptions of code sources for clustering.
\paragraph{Combining clustering with sample-and-rerank.}
The simplest approach is to refactor each cluster independently and take the union (concatenation) of each cluster's library:
\begin{align}
\library^\star &= 
\bigcup_{c\in \textsc{Cluster}_S(\left\{ \program_n \right\})} \library_c\\
\left\{ \program'_n \right\} &= 
\bigcup_{c\in \textsc{Cluster}_S(\left\{ \program_n \right\})}  \left\{ \program'_{c,i}\right\}_{i=1}^{|c|}\\
\library_c, \left\{ \program'_{c,i} \right\}_{i=1}^{|c|}&=
\argmin_{\library, \left\{ \program'_{c,i} \right\}_{i=1}^{|c|}\in \samplek{c}} 
\loss{\library, \left\{ \program'_{c,i} \right\}_{i=1}^{|c|}}\text{, for each }c\in \textsc{Cluster}_S(\left\{ \program_n \right\})
\end{align}
The approach above ignores the fact that library abstractions discovered in one cluster might be useful in another cluster.
A more sophisticated approach accumulates a library across clusters, and when refactoring a cluster, adds the accumulated library to the prompt.
This lets abstractions discovered earlier carry forward across the collection.
\Cref{sec:incrementallibrarian} describes this extension.

\section{What Makes a Good Refactoring?}
\label{sec:metric}

We compare different metrics $M$ measuring the quality of a refactoring:


\textbf{Tokens} measures the total number of tokens in the refactored \ITEM{}s \emph{and} in the library.
It minimizes program size, but not at the expense of creating a bloated library:
Simply replacing every program with its own monolithic library function would not improve the tokens metric, because it measures library size as well.
Concretely, $M_\text{tokens}(\mathcal{L},\{\rho'_n\}) = \textsc{Tokens}(\mathcal{L}) + \sum_{n} \textsc{Tokens}(\rho'_{n})$.

\textbf{Minimum Description Length (MDL)}
evaluates the negative log probability under a reference distribution, taking into account both the library and refactored sources.
Concretely, $M_\text{MDL}(\mathcal{L},\{\rho'_n\}) = - \log p_{\text{LM}}(\mathcal{L}) + \sum_{n} - \log p_{\text{LM}}(\rho'_{n} \mid \mathcal{L})$, where $p_{LM}(\rho_n'|\mathcal{L})$ is concatenating the library and the program into one prompt, but only counting the perplexity of the later program tokens.
This has a Bayesian justification: The MDL library is the maximum aposteriori estimate of $\mathcal{L}$ given conditionally-independent programs.
We use Qwen-2.5-3B  as our reference language model, as it is modern, open, and has  publicly-available endpoints for querying logits, which is required for scoring refactorings. To confirm that our MDL optimization results are not model specific, we computed MDL values for 15 Code Contests clusters, each of 50 valid refactoring candidates, using Qwen and Llama-3.2-3B models. We found a 94\% agreement in the minimum MDL candidates from both models.

\textbf{Cyclomatic Complexity (CC)} is a longstanding metric from the software engineering community which measures the number of linearly independent paths through a program's control flow graph~\citep{1702388}.
Smaller programs often have lower cyclomatic complexity.
It is equivalent to defining $M_\text{CC}(\mathcal{L},\{\rho'_n\}) = \textsc{cc}(\mathcal{L}) + \sum_{n} \textsc{cc}(\rho'_{n} )$ and $\textsc{cc}(\rho)=E(\rho)-N(\rho)+2P(\rho)$, where $E$, $N$, and $P$ measure the number of control flow edges, nodes, and connected components, respectively.

\textbf{Maintainability Index (MI)} is a modern software engineering metric combining several other metrics, including lines-of-code, cyclomatic complexity, and Halstead volume into a single score. Higher MI values are intended to indicate easier-to-maintain code, so we define $M_\text{MI}(\mathcal{L},\{\rho'_n\}) = -\textsc{MI}(\mathcal{L}) + \sum_{n} -\textsc{MI}(\rho'_{n} )$.


\subsection{Asymptotic behavior of metrics in large-sample regime}
\label{sec:which_objective}

\begin{figure}[h]
\includegraphics[width = \textwidth]{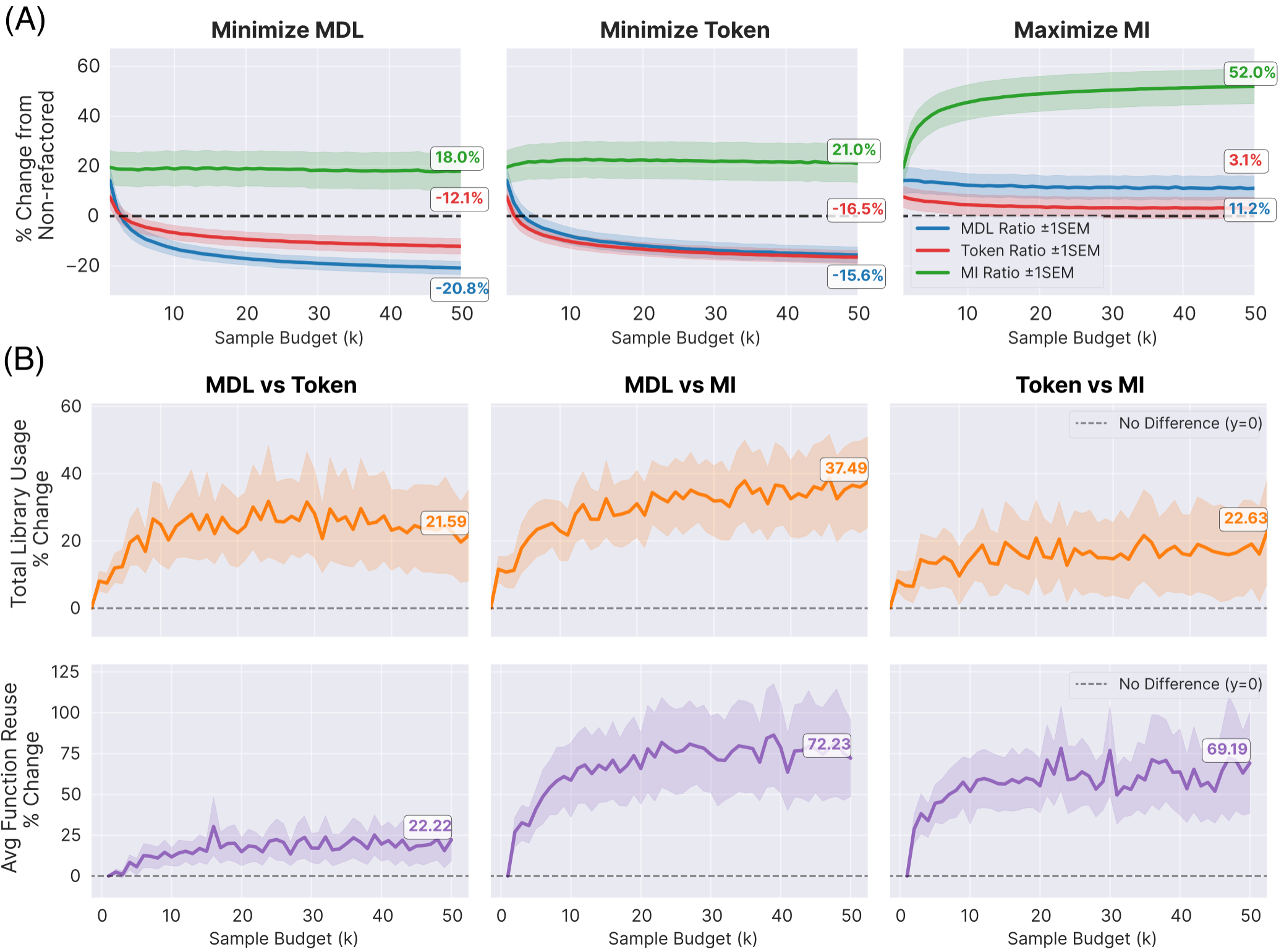}
\vspace{-15pt}
\caption{(A) Asymptotic behavior of metrics for scoring libraries and refactorings (columns) varying refactoring budget (horizontal axes).
(B) Comparing metrics via proxies of downstream library quality (total library usage and average calls per library function), for which MDL$>$Tokens$>$MI. All results are estimated using Best@k.
See also Appendix \Cref{fig:appendix_asymptotics}.
}
\label{fig:objectivecomparison}
\end{figure}

Are these metrics equally effective at encouraging modular and reusable libraries?
To answer this question, we run \NAME{} on 15 CodeContests  (each of three files) using MDL, tokens, maintainability index, and cyclomatic complexity, while varying the inference-time sample budget $K$ (\Cref{fig:objectivecomparison},\ref{fig:appendix_asymptotics}). We use Best@k estimator for the expected value of metrics for all $k \leq K$ (we describe the estimator and prove its correctness in Appendix \ref{ref:proof}).
Tokens and MDL separate cleanly from classic software engineering metrics:
Optimizing tokens/MDL, both of which essentially compress the original programs, does \emph{not} yield steady improvements in MI/CC, and vice-versa.
To understand whether these libraries expose shared abstractions, we examine the average number of times that each library routine is used, and the average number of library invocations per library function.
This teases apart tokens and MDL:
Optimizing MDL yields more more reusable libraries (used about $8\times$ per task), with each function called more often (called about $2.2\times$ per function)---exceeding the other metrics we consider.
See Appendix \Cref{fig:appendix_asymptotics} for full results, including Cyclomatic Complexity.

\begin{wrapfigure}{r}{0.4\textwidth}
    \centering
    \vspace{-20pt} 
    \includegraphics[width=0.4\textwidth]{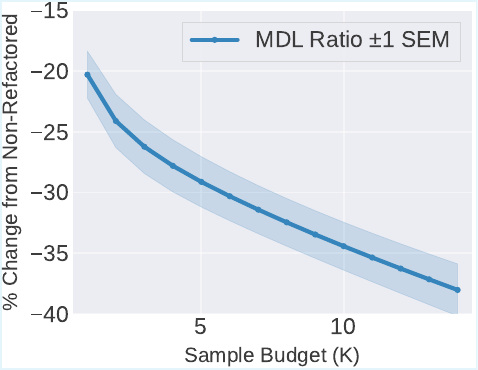}
    \vspace{-10pt} 
    \caption{Best@K MDL ratio. Increasing sample budget improves MDL on Transformers. }
    \label{fig:transformers_scaling}
    \vspace{-10pt} 
\end{wrapfigure}
Studying these metrics at large $k$ allows understanding their inference-time scaling behavior.
While the underlying metric itself improves (with diminishing returns), this was not to be taken for granted:
The backend language model must produce sufficient diversity to steadily improve these metrics.
Prior state-of-the-art, such as~\cite{10.5555/3692070.3693967}, instead take a single sample, but the results here suggest benefit from further test-time search, and indeed,
real-world repos benefit from steady improvement with increased samples (Figure~\ref{fig:transformers_scaling}).
But our proxies of library utility plateau much earlier, around $k=20$ samples, suggesting large $k$ is unnecessary in practice:
Effective library building benefits from test-time compute, but does not demand an exorbitant amount of it.

\newpage




\subsection{Human evaluation of refactoring metrics}
\begin{wrapfigure}{r}{0.45\textwidth}
    \centering
    \vspace{-10pt} 
    \includegraphics[width=0.45\textwidth]{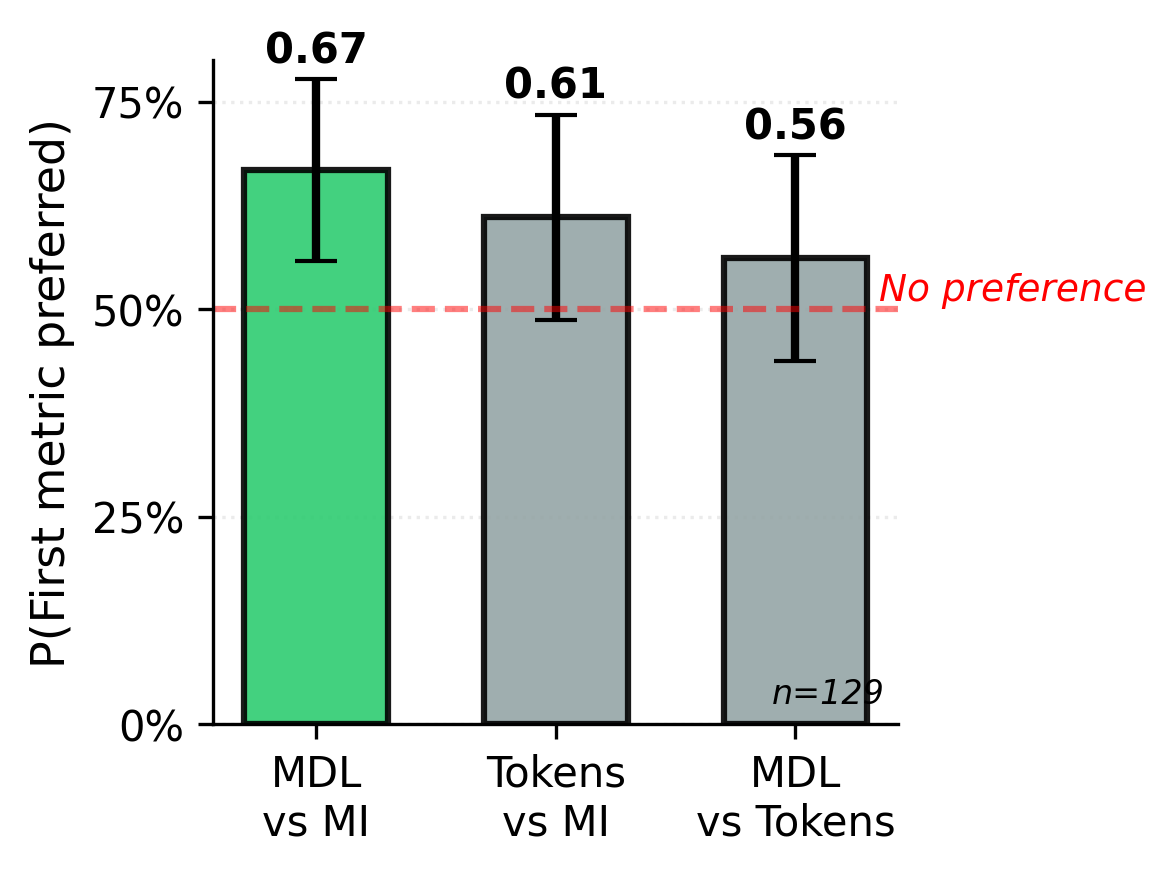}
    \vspace{-17pt}
    \caption{Human evaluation of different refactoring objectives. 
    Judges compare pairs of refactorings that both pass all test cases. 
    MDL aligns best with human preferences.}
    \label{fig:human}
    \vspace{-10pt} 
\end{wrapfigure}

We perform a human study to corroborate the findings of \Cref{sec:which_objective} using the exact same CodeContests clusters (\Cref{appendix:human_study}).
The human study
compares tokens, MDL, and Maintainability Index by (1) refactoring clusters into libraries, (2) presenting human participants with the original sources and their refactorings under pairs of  metrics, and (3) eliciting pairwise preferences from human participants.

Humans prefer MDL-minimizing libraries (\Cref{fig:human}), and although the preference is only statistically significant for MDL vs. MI, the  data suggest a rank-order preference of MDL>Tokens>MI; Tokens vs. MI was slightly below the threshold of significance.
We ran $N=14$ participants (eliciting 129 judgments), and although more participants would 
make more of these comparisons statistically significant, already we see a general preference for compression-based metrics (MDL and Tokens) with only MDL crossing the threshold of statistical significance.
Indeed, minimizing MDL is not quite the same as minimizing tokens, as
\Cref{fig:obfuscation-ltx} illustrates:
We believe basically every human coder would prefer the MDL-minimizing program in this example.

\textbf{We therefore adopt $M_{MDL}$ as the primary objective in the remainder of this paper:}
In addition to support from this human study, (1)  Bayesian arguments support MDL; (2) corner cases in the style of `Perl golf' provide existence proofs of the liability of merely minimizing tokens; and (3) reasonable proxies for library reuse favor MDL (\Cref{sec:which_objective}).

\begin{figure}[h]
\centering
\begin{minipage}[t]{0.5\textwidth}
\vspace{0pt}
\begin{moderncode}
from ..shared_library import (
    rotate_half,
    apply_rotary_pos_emb,
    repeat_kv,
    eager_attention_forward,
    RMSNorm,
    BaseMLP,
    BaseRotaryEmbedding,
    BaseAttention,
    BaseDecoderLayer,
)

class LlamaRMSNorm(RMSNorm):
    ...

class LlamaRotaryEmbedding(BaseRotaryEmbedding):
    ...

class LlamaMLP(BaseMLP):
    def __init__(self, config):
        super().__init__(
            config, 
            mlp_bias=config.mlp_bias
        )

class LlamaAttention(BaseAttention):
    def __init__(self, config: LlamaConfig, 
                 layer_idx: int):
        super().__init__(
            config=config,
            layer_idx=layer_idx,
            attn_bias=config.attention_bias,
            sliding_window=None
        )
...
\end{moderncode}
\end{minipage}
\hfill
\begin{minipage}[t]{0.46\textwidth}
\vspace{0pt}
\begin{moderncode}
from ..shared_library import rotate_half,apply_rotary_pos_emb,repeat_kv,eager_attention_forward,RMSNorm as R,BaseMLP as M,BaseRotaryEmbedding as E,BaseAttention as A,BaseDecoderLayer as Y
class Z(R):...
class I(E):...
class J(M):
 def __init__(s,g):
  super().__init__(g,mlp_bias=g.mlp_bias)
class H(A):
 def __init__(s,g:F,i:int):
  super().__init__(
   config=g,layer_idx=i,
   attn_bias=g.attention_bias,
   sliding_window=None)
class V(Y):
 def __init__(s,g:F,i:int):
  super().__init__(
   config=g,layer_idx=i,
   norm_class=Z,
   mlp_class=J,
   attention_class=H)
...
\end{moderncode}
\end{minipage}
\caption{Example where tokens and MDL diverge:
Obfuscating the original library definitions (left) by shortening variable names  (right) reduces tokens but increases MDL.
}
\vspace{-10pt}
\label{fig:obfuscation-ltx}
\end{figure}

\section{What we learn from running \NAME{} on \BENCH{}}

We empirically study \NAME{} on \BENCH{} with the goal of understanding (1) the degree to which library abstractions are reused across programs, (2) how our method compares to state-of-the-art library learning on existing datasets, and 
(3) whether \NAME{} holds value for real-world repos.


\paragraph{\NAME{} discovers reusable functions for competition programming--but some functions are only called once.}
We test on CodeContests with a cluster size of $S=3$ and a sample budget of $K=8$ draws from o4-mini, as reasoning models perform well on competition programming.\footnote{Code agents such as Codex, Claude Code, and others underperformed o4-mini (\Cref{app:results})}
\Cref{tbl:librarian-metrics} shows that the resulting refactors and libraries approximately halve the MDL, which incidentally reduces program size as well (44\% relative reduction in token count).
Pass rate modestly improves as an incidental consequence of sampling and filtering with test cases.
Libraries average 10 functions, each heavily reused:
Averaging 5 uses per function within tasks comprising only 10 programs.
But almost 40\% of library functions are only used once.
Why is that?

A signature of the MDL objective is a preference for whatever a language model assigns high apriori probability to.
Although a single-use function does not reduce line count or tokens ---the function could simply be inlined---it improves MDL if it yields a more natural decomposition of the target programs.
Indeed, human-written libraries sometimes include functions that are seldom used, provided they serve as a conceptually modular abstraction.
We therefore see single-use functions as a feature, not a bug. See Appendix \ref{ref:cc_example} for an example refactoring candidate on CodeContests.

\paragraph{Are these libraries useful for solving new, unseen programming problems?} 

Library learning has long sought to learn libraries from training programs which then help solve new unseen program synthesis tasks.
The Logo and Date datasets fit within this paradigm.
Recently REGAL improved the state-of-the-art on these library learning datasets.
Because our clustering is heavily inspired by REGAL, for fair comparison, we keep exactly their clustering setup but add MDL-based reranking using $K=5$ samples.
Despite the simplicity of these datasets, we find value in our more complicated method.
\Cref{tbl:regal-datasets} shows that sampling and reranking by MDL yields up to a 41.8\% relative improvement in solve rate on unseen programming problems, and that even when the gains are more modest, we still improve upon the state-of-the-art.
But these are relatively simple problems solvable with about ten lines of code---does this work in the real world?

\begin{table}[H]
  \centering
  \begin{minipage}[t]{0.4\textwidth}
    \centering
    \small
    \captionof{table}{Results for \NAME{} on 10 Code Contests tasks ($K=8,S=3$)}
    \label{tbl:librarian-metrics}
    \begin{tabular}{lr}
    \toprule
    \textbf{Metric} & \textbf{Value} \\
    \midrule
    Pass Rate & 90.67\% ±1.88 \\
    Pass Rate Improvement & 6.33\% ±1.41 \\
    MDL Ratio & 0.53 ±0.03 \\
    Token Ratio & 0.66 ±0.04 \\
    Library Functions & 10.30 ±1.41 \\
    Avg Calls per Function & 5.17 ±1.08 \\
    \% Single Use Functions & 38.03\% ±4.88 \\
    \bottomrule
    \end{tabular}
  \end{minipage}\hfill
  \begin{minipage}[t]{0.5\textwidth}
    \centering
    \small
    \captionof{table}{Solving holdout test program synthesis tasks using learned libraries.}
    \label{tbl:regal-datasets}
    \begin{tabular}{llr}
    \toprule
    \textbf{Dataset} & \textbf{Model} & \textbf{Pass Rate} \\
    \midrule
    \multirow{2}{*}{Logo} 
    & REGAL (gpt-3.5-turbo) & 49.3\% ±1.1 \\
    & \NAME{} (3.5-turbo)   & 69.9\% ±0.9 \\
    \midrule
    \multirow{2}{*}{Date} 
    & REGAL (gpt-3.5-turbo) & 90.2\% ±0.5 \\
    & \NAME{} (3.5-turbo)   & 94.7\% ±0.7 \\
    \bottomrule
    \end{tabular}
  \end{minipage}
  
\end{table}

\textbf{Real-World Refactoring.} 
The HuggingFace Transformers library is used by nearly 400k GitHub projects.
We deploy \NAME{} to 10 source files, using Claude Code to sample $K=15$ refactorings per cluster of size $S=5$, believing an agent such as Claude Code would excel at repo-level edits.
\NAME{} distilled repeated abstractions such as MLPs, Attention, Decoder classes, RoPE helper functions, etc., lowering MDL to 67.2\% of its original value while still passing all integration tests. The top-3 refactorings based on MDL have an average of $18 \pm 4.4$ abstractions (functions, classes) in the library, each of which is called on average $4.59 \pm 0.39$ times in the refactored models.
For Diffusers, scheduler clusters yielded top-3 MDL refactorings with an average of $12.3 \pm 1.6$ functions and $3.0 \pm 0.4$ calls per function, while UNet refactorings produced richer abstractions with an average of $17.0 \pm 5.6$ functions/classes and $3.43 \pm 0.67$ calls each.

\begin{figure}[h    t]
    \centering
    \includegraphics[width=\linewidth]{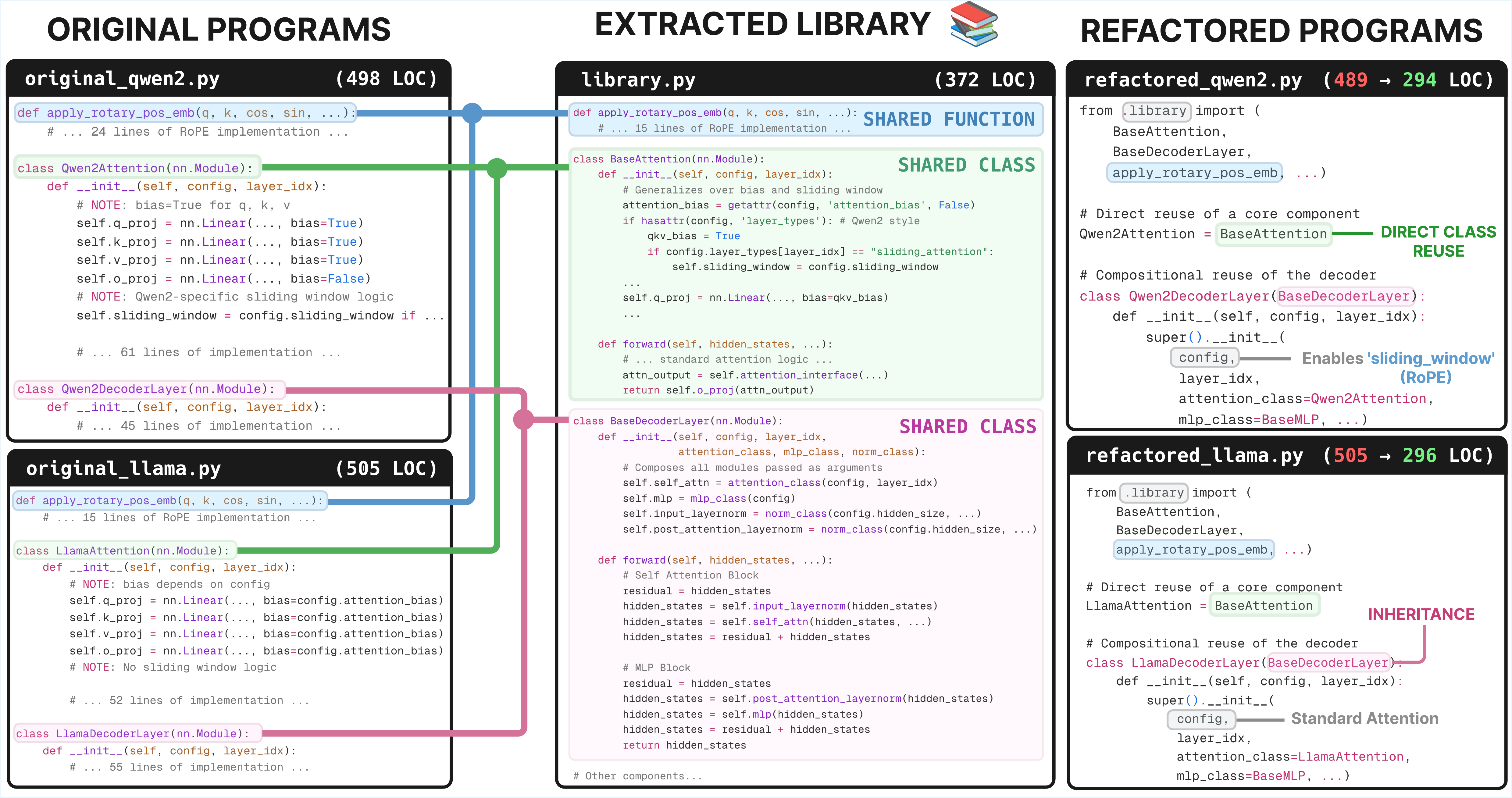}
    \vspace{-10pt}
    \caption{Representative result for refactoring HuggingFace  Transformers using \NAME{}}
    \label{fig:transformersillustration}
    \vspace{-5pt}
\end{figure}

Refactoring at scale proved expensive:
Each refactoring took approximately 30 minutes to generate and test.
But this is a one-off cost, and in our view, the refactored Transformers and Diffusers sources are much cleaner, and the new library is transparently reusable (\Cref{fig:transformersillustration}).
To the best of our knowledge, this is the first time any library learning algorithm has been successfully applied to real-world software projects.

\textbf{Learned libraries from these real-world codebases are useful for unseen downstream refactoring tasks.} When a library learned on one cluster of Transformer files (5 models) is applied to refactor a second cluster, \NAME{} reduces the unseen cluster's MDL to 73\% of the its original value, with an average of 3.0 calls per library function. This demonstrates that \NAME{} learned libraries that can be repurposed to more compactly rewrite unseen real-world code sources.

\textbf{Human performance}. HuggingFace is concurrently refactoring the Transformers library. On the same 10 source files used above, their refactoring achieves an MDL compression ratio of 66.5\%. To estimate an independent ceiling, we the authors refactor the task ourselves and obtain 62\%. This suggests room for a little improvement beyond \NAME{}’s 67.2\%, but not by much (from inspection, specifically with better implementations of mixture-of-experts and vision models).

\section{Conclusion}
We introduce a new benchmark \BENCH{} and method \NAME{} for compressing \ITEM{}s through reusable abstractions.
We highlight the challenges of producing modular and maintainable libraries, then present an effective method for using LLMs to do this task.
By framing refactoring as an optimization problem, our work opens new directions for building more general and scalable code understanding and generation systems.
In particular, the structure of \BENCH{} lends itself well to reinforcement learning, where training would entail synthesizing collections of repositories to refactor then computing rewards based on MDL or other metrics.

\textbf{Limitations.}
We evaluate on synthetic toy problems (Logo and Date), and on competition programming problems, neither of which are naturalistic, although this is partly counterbalanced by our study of real-world refactoring.
Compression, whether measured by tokens or MDL, may not always correlate with reuse, a limitation we sought to address through
our experiments on downstream programming problems, and on holdout Transformers files:
But investigating reuse on unseen app-building problems for real-world repo-level refactors remains open.


\section*{Acknowledgements}
KE and ZC were supported by NSF grant \#2310350.

\bibliography{references}
\bibliographystyle{iclr2026_conference}
\clearpage


\appendix

\section{Incremental Version of \NAME{}}\label{sec:incrementallibrarian}
We support sharing library components across clusters through the below incremental version of our algorithm.
It processes clusters sequentially, rather than in parallel, and its output can depend on the ordering imposed upon the clusters.
\begin{gather*}
\mathcal{L}_0 = \varnothing \\[6pt]
\{\rho_n'\} = \bigcup_{c \in \text{CLUSTER}(\{\rho_n\})} \{\rho_{c,i}'\}_{i=1}^{|c|} \\[6pt]
(\Delta\mathcal{L}_c, \{\rho_{c,i}'\}_{i=1}^{|c|})
= \arg\min_{(\Delta\mathcal{L}, \{\rho_i'\}) \in \text{SAMPLE}_k(c;\,\mathcal{L}_{t-1})}
\ \ell(\mathcal{L}_{t-1}\cup\Delta\mathcal{L}, \{\rho_i'\}_{i=1}^{|c|})
\quad \forall c \in \text{CLUSTER}(\{\rho_n\}) \\[6pt]
\mathcal{L}_t = \mathcal{L}_{t-1} \cup \Delta\mathcal{L}_t,\qquad
\mathcal{L}^\star = \mathcal{L}_{|\text{CLUSTER}(\{\rho_n\})|}
\end{gather*}

\section{Algorithm}
\label{app:algo}

\begin{algorithm}
\caption{Refactoring Specialized Programs into a Joint Library}
\label{alg:refactoring}
\begin{algorithmic}[1]
\Require Set of independent, specialized programs $P_{initial} = \{\program_1, \program_2, \dots, \program_n\}$
\Require Sample Budget $K$
\Ensure Joint library $\library_{final}$ and set of refactored programs $P_{final}$
\State $\mathcal{C} \gets \text{Cluster}(P_{initial})$
\State $\library_{final} \gets \emptyset$, $P_{final} \gets \emptyset$

\ForAll{cluster $c \in \mathcal{C}$ } \Comment{Each cluster independently}
    \State $T_C \gets \text{GroupIntoTuples}(c)$ \Comment{Get tuples for each cluster}
    \ForAll {tuples $\tau \in T_C$}
        \State $\{f_{retrieved}\} \gets \text{RetrieveRelevantFromLibrary}(\library, \tau)$
        \State $S \gets \emptyset$
         \For{$i = 1 \text{ to } K$} \Comment{Sample $k$ times}
            \State $(\{f_{new,i}\}, \{\rho'_i\}\}) \gets \sample{f_{retrieved}, \tau}$ 
            \State $S \gets S \cup \{(\{f_{new,i}\},\{\rho'_{i}\} )\}$
        \EndFor
        \State $(f_{best}, \{\rho'_{best}\}) \gets \text{RerankAndSelectBest}(S, \loss{\cdot})$ \Comment{Rerank using objective}
        \State $\library_{final} \cup \{f_{best}\}$
        \State $ P_{final} \cup \{\rho'_{best}\}$ 
    \EndFor    
\EndFor

\State \Return $\library_{final}, P_{final}$

\end{algorithmic}
\end{algorithm}





\section{Experimental Setup}

\paragraph{Grouping Programs into Collections}
To facilitate parallel application of \NAME{} and manage the dataset scale, we assume that semantically distant \ITEM{}s will have minimal overlap in their optimal library functions. Therefore, our overall approach partitions the dataset into disjoint collections through clustering.

For \textbf{CodeContests}, these collections are constructed from an initial corpus of $\sim$9k problems with Python solutions: We first filter these \ITEM{}s, removing those whose selected canonical solution is under 10 lines (minimal refactoring potential). For the remaining 4596 solutions we use a language model to generate textual descriptions of canonical solutions—emphasizing reusable components—which are embedded using OpenAI's \texttt{text-embedding-ada-002}. 

Agglomerative Clustering \citep{ward1963hierarchical} is subsequently applied to these embeddings to partition the \ITEM s\ into a predefined number of initial clusters, in our case 120. To create uniformly sized experimental units, we subsample each such cluster to form collections of 30 \ITEM s. This collection size was empirically chosen because it balanced between the runtime of \NAME{} without limiting compression. We select 10 collections that we then use to evaluate our methods.


For Transformers, since the number of models is on the lower end, we manually chose a set of popular LLM / VLM models and passed them to the agent in collections of 5 code sources.

\paragraph{REGAL Baselines.}
To evaluate the ability of our libraries to support reuse on new problems, we turn to the  program synthesis tasks used in REGAL, where learned libraries are added to help the program synthesizer. We evaluate on the two domains published by the authors, Logo and Date.
Because our clustering is inspired by REGAL but adds additional complexity, for fair comparison, we keep their setup the same and only augment the training using sample + MDL rerank procedure described in \Cref{sec:method}.

\paragraph{Code Contests.}
To evaluate \NAME \ on refactoring Code Contests we select 6 collections of 30 \ITEM s (problems). In each collection we group the problems into tuples of size $3$. We set the sample budget to be $K=8$, since our ablations show that with larger $K$ we  discover better libraries \ref{fig:objectivecomparison}. We use the MDL objective for rankings. 
The model used for sampling is OpenAI's o4-mini~\citep{openai2024o4mini}.
To obtain MDL scores we use Qwen 2.5 7B Instruct~\citep{qwen2025qwen25technicalreport}
as a balance between quality, speed, and cost.

\paragraph{Code Agents on Transformers and Diffusers Repositories.}

To fairly evaluate performance on the task by state-of-the-art systems, we use coding agents that advertise long-context ability to reason about, write, and refactor code repositories. Specifically, we use Claude Code (Cl)~\citep{claudecode} which uses the Opus 4.1.

We test whether code agents can refactor collections of code sources autonomously, without human intervention.
Refactoring repositories with code agents involves planning and iterative (re-)implementation and testing. Code agents are prompted to perform each of these steps, with feedback from the unit tests. Agents must run and repair unit tests autonomously.
We run coding agents multiple times per task, logging their progress in checklists stored in text files.


\newpage

The instruction provided to human evaluators is as follows:

\begin{figure}[H]
\begin{lstlisting}[style=codestyle, caption={Human Evaluation Instruction}, label={human_eval_prompt}]


## 1. Materials Provided

You will be given a set of files for each example case:

* **`original_programs.py`**: This file contains a set of 3 distinct Python programs, each presented with its corresponding problem description/query. This represents the "before" state.
* **`v1.py`**: This file presents the first refactoring approach. It includes:
    * The 3 refactored versions of the original programs.
    * A "library" section (e.g., `codebank.py` or inline) containing helper functions. These helper functions might be retrieved from an existing common library or newly created during this refactoring.
    * Either the retrieved or the new helper function sections may be _empty_, in case no programs existed in the codebank at the time or if no need helper functions were created by the LLM.
* **`v2.py`**: This file presents the second, alternative refactoring approach. Similar to `refactoring_v1.py`, it includes:
    * The 3 refactored versions of the original programs (using a different strategy than v1).
    * A "library" section with its own set of helper functions.


**NOTE**:  both refactorings had accuracy at least as good as the original programs.

## 2. Your Task

Your primary task is to:

1.  **Review** the `original_programs.py` to understand the initial code and the problems being solved.
2.  **Analyze** both `refactoring_v1.py` and `refactoring_v2.py`. Pay close attention to how the original programs have been restructured and what functionalities have been extracted into their respective libraries.
3.  **Decide which refactoring (Version 1 or Version 2) you believe is "better,"** based on the evaluation criteria provided below (or your own criteria!).

## 3. Evaluation Criteria: What to Consider for Your Choice

When comparing `refactoring_v1.py` and `refactoring_v2.py`, please *consider* the following aspects to inform your choice. The "better" refactoring should ideally excel in these areas:

> Most importantly, make sure that the extracted functions are **actually reusable and not too specific.** If the main programs are short, the refactoring is not immediately "better"! Try to think whether the extracted functions could actually be used in a different program down the line.

* **Reusability of Helper Functions :**
    * **Generality:** Are the new helper functions general-purpose and potentially useful for *other, different* programs and problems beyond the three presented?
    * **Reuse:** How much were existing helper functions reused?
    * **Specificity:** Are the functions too specialized to the current set of problems, limiting their broader applicability? _Avoid functions that are essentially just the original program broken out into a "helper."_
    * Composability
* **Maintainability:**
    * Readability & Understandability
    * Ease of Modification
    * Separation of Concerns

## 4. What NOT to Focus On:

* **Comments:** Please disregard the presence or absence of comments in the code for this evaluation. These are superficially generated by LLMs in some occasions and could be added manually after with a single pass.
* **Minor Stylistic Differences:** Do not focus on trivial differences in variable naming or formatting, unless they significantly impact readability or understanding.

## 5. How to Provide Your Feedback

For each example case, please provide:

1.  **Your Preferred Version:** (e.g., "Version 1" or "Version 2")
\end{lstlisting}
\end{figure}

\section{Further Results on Refactoring Metrics}
Here we include the full graph of asymptotic behavior of all four scoring metrics (MDL, tokens, MI and CC), reporting the raw library metrics and the metric ratios. We can see that refactored programs have higher CC than the baseline and that optimizing CC as an objective does not decrese the other metrics.
MDL performs best on total raw library usage as well as on average times a library function is used in the refactorings. MI ends up optimizing the number of library functions the best, but their reusability is below the average for the refactorings. Optimizing for tokens produces smaller libraries with less usage per function compared to optimizing MDL.

\begin{figure}
    \centering
    \includegraphics[width=\linewidth]{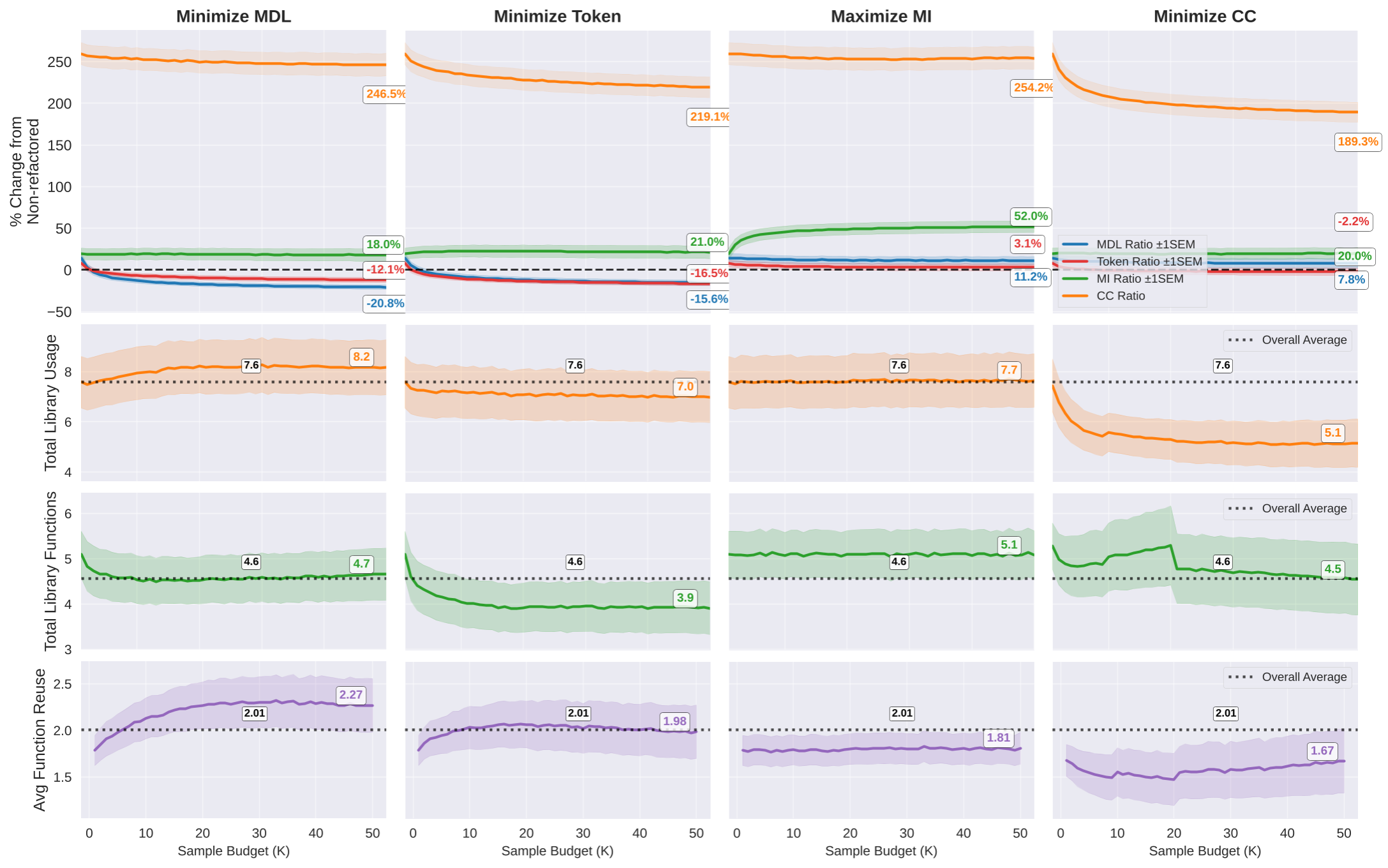}
    \caption{Asymptotic behavior of metrics for scoring libraries and refactorings (columns) varying refactoring budget (horizontal axes).}
    \label{fig:appendix_asymptotics}
\end{figure}

\section{Human Study Details}
\label{appendix:human_study}
We ran a user study with 14 participants where each participant had to judge 10 refactoring pairs. Each pair was comparing two out of three metrics, MDL, tokens, and MI. We showed the participants the original programs, as well as both of the refactoring versions (including the programs and the learned libraries). The participants were able to choose either version or say that the two refactorings are almost the same.

To quantify pairwise preferences between refactoring metrics, we employed the Bradley-Terry model, a standard framework for analyzing paired comparison data. We fit the model using maximum likelihood estimation with mutual information (MI) as the reference category $(\pi_MI = 1)$. To address potential noise in human judgments, we applied consensus-based filtering with a 75\% threshold, retaining all responses for comparisons with low consensus (indicating genuine ambiguity) while excluding minority responses on high-consensus comparisons where the majority preference likely indicates the correct judgment. This conservative approach preserved 92.4\% of responses while strengthening the statistical evidence for metric preferences, with MDL significantly outperforming MI ($p = 0.67$, $95\%$ Confidence Interval: $[0.56, 0.78]$).
Even without the filtering MDL preference over MI was statistically significant.


\section{Best@k Compression is a U-Statistic}
\label{ref:proof}

We wish to estimate the expected compression ratio achieved by our \emph{sample + rerank} method, which samples $k$ candidate refactorings, discards any that do not pass the tests, and selects the one with the lowest score (total log-prob).

\textbf{Background on U-Statistics. }Let $Z_1,\dots,Z_n \overset{\text{i.i.d.}}{\sim} F$. For a symmetric function $h: \mathcal{Z}^k \to \mathbb{R}$, the \emph{U-statistic of order $k$} is defined as
\begin{equation}
    U_n \;=\; \binom{n}{k}^{-1} \sum_{1 \le i_1 < \cdots < i_k \le n} h(Z_{i_1}, \dots, Z_{i_k}).
\end{equation}
By construction,
\begin{equation}
    \mathbb{E}[U_n] \;=\; \mathbb{E}[h(Z_1,\dots,Z_k)],
\end{equation}
so $U_n$ is an unbiased estimator of the population quantity $\theta = \mathbb{E}[h(Z_1,\dots,Z_k)]$.

\textbf{Application to Best@k Compression.} Let each valid refactoring be a pair $Z = (S,C)$, where $S$ is the score and $C$ is the compression ratio. Define the symmetric function
\begin{equation}
    h_k(z_1,\dots,z_k) \;=\; C_{j^*}, 
    \qquad j^* = \arg\min_{1 \le j \le k} S_j,
\end{equation}
the compression ratio of the lowest-score refactoring among $k$ draws.
The population target is then
\begin{equation}
    \theta_k = \mathbb{E}[h_k(Z_1,\dots,Z_k)].
\end{equation}

Given $n$ valid samples, our estimator is
\begin{equation}
    \widehat{\theta}_k 
    \;=\; \binom{n}{k}^{-1} 
    \sum_{1 \le i_1 < \cdots < i_k \le n} 
    h_k(Z_{i_1}, \dots, Z_{i_k}).
\end{equation}

\noindent\textbf{Proposition.} 
$\widehat{\theta}_k$ is a U-statistic of order $k$ with function $h_k$, and hence an unbiased estimator of $\theta_k$.

\paragraph{Proof.}
(1) \emph{Symmetry of the function $h_k$.}  
$h_k$ selects the compression associated with the lowest score among its $k$ arguments. Permuting the inputs does not affect this outcome (ties can be resolved with a fixed, permutation-invariant rule). Thus $h_k$ is symmetric.

(2) \emph{U-statistic form.}  
By definition, a U-statistic of order $k$ with kernel $h_k$ is
\[
U_n = \binom{n}{k}^{-1} \sum_{1 \le i_1 < \cdots < i_k \le n} h_k(Z_{i_1}, \dots, Z_{i_k}),
\]
which matches $\widehat{\theta}_k$ exactly.

Therefore, $\widehat{\theta}_k$ is a U-statistic of order $k$. By the unbiasedness property of U-statistics,
\[
\mathbb{E}[\widehat{\theta}_k] = \theta_k.
\]
\hfill$\square$

\medskip
Thus, our reported \emph{best@k compression curves} provide unbiased estimates of the expected performance of the \emph{sample + rerank} method.

\section{Clustering Analysis: CodeContests} 
\label{sec:analysis_clusters}
\begin{figure}[h]
\includegraphics[width = \textwidth]{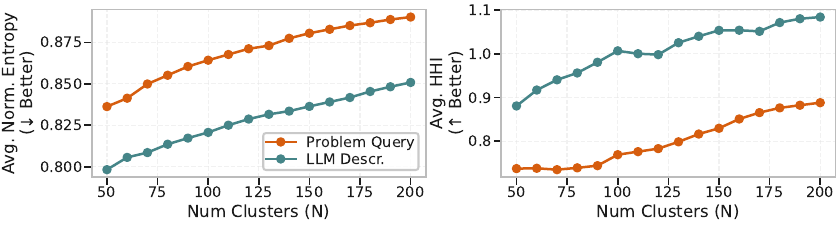}
\caption{Clustering analysis of 4,596 Code Contest problems, comparing the thematic coherence of clusters formed using our proposed method 
versus REGAL-style clustering.
}\label{fig:clustering_analysis}
\end{figure}

We analyze the coherence of the clusters underlying collections in \BENCH{}-\textbf{CodeContests}.
In particular, we compare clustering based on o4-mini generated \ITEM{} descriptions against task descriptions.
Since task descriptions in competition coding problems are designed to hide the algorithmic approach needed to solve problem, we expect that clusters based on \ITEM{} descriptions are more coherent. We use Normalized Tag Instance Entropy and   Herfindahl-Hirschman Index to evaluate clusterings. \Cref{fig:clustering_analysis} shows our clustering approach yields more thematically coherent clusters, evidenced by achieving lower entropy and higher HHI values across the entire tested range of $N$. We provide definitions of our measures below.




\subsection{Collection Coherence Measures}
\label{app:coherence}
We use two measures
to evaluate the thematic coherence of collections:
Good collections should group \ITEM{}s with a (1) concentrated and (2) identifiable set of shared \emph{conceptual tags}, which for CodeContests are provided as ground truth (\texttt{trees}, \texttt{graphs}, 
etc.).

We provide the full definitions of the collection coherence measures  below.

\paragraph{Normalized Tag Instance Entropy:}
This measures the concentration of tag \textit{instances} within a collection $C$.
Let $p_i$ be the proportion of the $i$-th unique tag type among all tag instances in $C$, and $D_C$ be the number of distinct tag types in $C$. If $D_C > 1$, the normalized entropy $H_N$ is defined as:
\begin{equation}
H_N = - \frac{\sum_{i=1}^{D_C} p_i \log_2 p_i}{\log_2 D_C}
\end{equation}
If $D_C \le 1$, then $H_N=0$. Lower $H_N$ (closer to 0) indicates higher thematic purity, meaning fewer tag types dominate the bulk of tag mentions.

\paragraph{Herfindahl-Hirschman Index (HHI) for Problem Presence:}
This measures tag concentration across distinct \textit{problems} in a cluster $C$. Let $s_t$ be the proportion of problems in $C$ that include tag $t$ (a problem contributes to $s_t$ if $t$ is one of its unique tags). A higher HHI signifies that the problems are collectively characterized by a smaller, more focused set of tags.
\begin{equation}
\text{HHI} = \sum_{t \in \text{Tags}(C)} s_t^2
\end{equation}
where $\text{Tags}(C)$ represents the set of unique tags present in cluster $C$.

\section{Benchmark Comparison}
We compare our benchmark, \BENCH{}, to similar benchmarks in Table \ref{tab:benchmark_comparison}.
We define creativity and design as the need to explore diverse solutions in order to find the best solution possible.
For example, optimizing for program correctness alone does not require exploring a large solutions space, whereas optimizing a program for speed would.
In the case of compressing large \ITEM{}s, we must explore the large space of shared abstractions afforded by libraries in order to maximize compression.

\begin{table}[H]
\centering
\caption{Comparison of Code Benchmarks}
\begin{tabular}{lcc}
\toprule
\textbf{Benchmark} & \textbf{Creativity/Design} & \textbf{Scale} \\
\midrule
SWE-bench \citep{jimenez2024swebench} & Low & Repository \\
Commit-0 \citep{zhao2025commit} & Medium & Repository \\
RefactorBench \citep{gautam2025refactorbench} & Low & File \\
ECCO \citep{waghjale2024ecco} & High & Function \\
KernelBench \citep{ouyang2025kernelbenchllmswriteefficient} & High & Function \\
\BENCH (Ours) & High & Repository \\
\bottomrule
\end{tabular}
\label{tab:benchmark_comparison}
\end{table}

\section{Full \BENCH{} Codecontests Results}
\label{app:results}
We present the full agent scores for the CodeContests split in Table \ref{tbl:codecontests-results}. The results are given both for each cluster of code sources, as well as averaged across clusters.

\begin{table}[H]
\centering
\begin{tabular}{@{}clrrrrr@{}}
\toprule
\multicolumn{1}{c}{\textbf{Cluster}} & \multicolumn{1}{c}{\textbf{Agent}} & \multicolumn{1}{c}{\textbf{Tokens}} & \multicolumn{1}{c}{\textbf{CC}} & \multicolumn{1}{c}{\textbf{Pass \%}} & \multicolumn{1}{c}{\textbf{MDL}} & \multicolumn{1}{c}{\textbf{MDL \%}} \\
\midrule
\multirow{4}{*}{0} & original & 9088 & 95 & 80.3 & 11745.85 & 100.0 \\
& sonnet 3.7 & 18114 & 176 & 87.0 & 15005.18 & 127.7 \\
& sonnet 4 & 11121 & 138 & 80.3 & 9901.53 & 84.3 \\
& codex-mini & 9321 & 95 & 80.3 & 9990.74 & 85.1 \\
\midrule
\multirow{4}{*}{1} & original & 12531 & 255 & 89.7 & 13431.86 & 100.0 \\
& sonnet 3.7 & 10470 & 239 & 96.7 & 8933.65 & 66.5 \\
& sonnet 4 & 11325 & 298 & 96.7 & 8214.42 & 61.2 \\
& codex-mini & 12762 & 255 & 89.7 & 11798.73 & 87.8 \\
\midrule
\multirow{4}{*}{2} & original & 14087 & 376 & 89.0 & 15012.77 & 100.0 \\
& sonnet 3.7 & 17345 & 429 & 91.3 & 13145.02 & 87.6 \\
& sonnet 4 & 14270 & 356 & 93.0 & 10522.66 & 70.1 \\
& codex-mini & 14318 & 376 & 89.0 & 13273.81 & 88.4 \\
\midrule
\multirow{4}{*}{3} & original & 14261 & 246 & 90.3 & 13348.82 & 100.0 \\
& sonnet 3.7 & 20749 & 241 & 97.7 & 15859.02 & 118.8 \\
& sonnet 4 & 13433 & 197 & 80.7 & 11937.04 & 89.4 \\
& codex-mini & 14495 & 246 & 90.3 & 11616.41 & 87.0 \\
\midrule
\multirow{4}{*}{4} & original & 17693 & 336 & 80.7 & 14665.16 & 100.0 \\
& sonnet 3.7 & 29860 & 358 & 100.0 & 20666.52 & 141.0 \\
& sonnet 4 & 18684 & 352 & 82.0 & 12801.21 & 87.3 \\
& codex-mini & 17923 & 336 & 80.7 & 12902.09 & 88.0 \\
\midrule
\multirow{4}{*}{5} & original & 12588 & 286 & 92.0 & 12790.11 & 100.0 \\
& sonnet 3.7 & 10580 & 128 & 99.3 & 8435.12 & 65.9 \\
& sonnet 4 & 10416 & 155 & 99.3 & 9167.85 & 71.7 \\
& codex-mini & 12819 & 286 & 92.0 & 11086.19 & 86.7 \\
\midrule
\multirow{4}{*}{6} & original & 11020 & 131 & 54.3 & 13540.41 & 100.0 \\
& sonnet 3.7 & 21747 & 502 & 88.0 & 19446.07 & 143.6 \\
& sonnet 4 & 11177 & 143 & 57.3 & 10492.00 & 77.5 \\
& codex-mini & 11251 & 131 & 54.3 & 11651.65 & 86.1 \\
\midrule
\multirow{4}{*}{7} & original & 12301 & 180 & 80.0 & 12393.73 & 100.0 \\
& sonnet 3.7 & 16390 & 166 & 91.0 & 13371.59 & 107.9 \\
& sonnet 4 & 11625 & 150 & 85.7 & 9304.25 & 75.1 \\
& codex-mini & 12534 & 180 & 80.0 & 10549.04 & 85.1 \\
\midrule
\multirow{4}{*}{Avg} & original & 12946 & 238 & 82.0 & 13366.09 & 100.0 \\
& sonnet 3.7 & 18157 & 280 & 93.9 & 14357.77 & 107.4 \\
& sonnet 4 & 12756 & 224 & 84.4 & 10292.62 & 77.1 \\
& codex-mini & 13178 & 238 & 82.0 & 11608.58 & 86.8 \\
\bottomrule
\end{tabular}
\caption{Comparison of the pass rate and compression metrics of the original files, Claude Sonnet 4 and codex-mini refactorings across CodeContests clusters.}
\label{tbl:codecontests-results}
\end{table}

\section{Refactoring examples of \NAME \ on Code Contests }
\label{ref:cc_example}
\subsection{Example 1}
In code snippets \ref{code:v1}, \ref{helper:v1}, \ref{code:v2}, \ref{helper:v2} one example of 2 refactoring versions. Specifically, the versions are both passing at least as many test cases as the original and they have the biggest difference in MDL among all the sample refactorings for that tuple. Sample + rerank filtering selected refactoring V2. You can observe that refactoring V1 introduces some problem specific functions like \texttt{build\_max\_beauty\_perm()}, while refactoring V2 sticks to more generally useful functions.

\begin{figure}[H]
\begin{lstlisting}[style=codestyle, caption={Version 1, New Helpers}, label={helper:v1}]
# ==== NEW HELPER FUNCTIONS ====
def compute_full_mask(i):
    """Return mask of all 1s of the bit-length of i."""
    return (1 << i.bit_length()) - 1

def build_max_beauty_perm(n):
    """Build permutation of 0..n maximizing sum of i^p[i]."""
    ans = [0] * (n + 1)
    used = set()
    for i in range(n, -1, -1):
        if i in used:
            continue
        mask = compute_full_mask(i)
        j = i ^ mask
        ans[i], ans[j] = j, i
        used.add(i)
        used.add(j)
    beauty = sum(i ^ ans[i] for i in range(n + 1))
    return ans, beauty

def solve_xor_sum(u, v):
    """
    Find shortest array whose xor is u and sum is v.
    Return list or None if impossible.
    """
    if u > v or (v - u) % 2:
        return None
    if u == v:
        return [] if u == 0 else [u]
    x = (v - u) // 2
    # try two elements
    if ((u + x) ^ x) == u:
        return [u + x, x]
    # fallback to three elements
    return [u, x, x]

def build_trie(keys):
    """
    Build a binary trie with counts for 30-bit numbers.
    Each node: [left_index, right_index, count].
    """
    tree = [[0, 0, 0]]
    for x in keys:
        now = 0
        tree[now][2] += 1
        for i in range(29, -1, -1):
            b = (x >> i) & 1
            if tree[now][b] == 0:
                tree[now][b] = len(tree)
                tree.append([0, 0, 0])
            now = tree[now][b]
            tree[now][2] += 1
    return tree

def trie_pop_min_xor(tree, x):
    """
    Pop one key from trie to minimize x^key and return that minimal xor.
    Decrements counts along the path.
    """
    now = 0
    res = 0
    for i in range(29, -1, -1):
        b = (x >> i) & 1
        nxt = tree[now][b]
        if nxt and tree[nxt][2] > 0:
            now = nxt
        else:
            now = tree[now][b ^ 1]
            res |= (1 << i)
        tree[now][2] -= 1
    return res
\end{lstlisting}
\end{figure}

\begin{figure}[H]
\begin{lstlisting}[style=codestyle, caption={Version 1, Refactored Programs}, label={code:v1}]

# ########## PROGRAM: node_16:cc_python_16 ##########

from codebank import *

def main():
    import sys
    data = sys.stdin.readline()
    if not data:
        return
    n = int(data)
    perm, beauty = build_max_beauty_perm(n)
    print(beauty)
    print(*perm)

if __name__ == "__main__":
    main()

# ########## PROGRAM: node_19:cc_python_19 ##########

from codebank import *

def main():
    import sys
    data = sys.stdin.readline
    n = int(data())
    A = list(map(int, data().split()))
    P = list(map(int, data().split()))
    trie = build_trie(P)
    O = [trie_pop_min_xor(trie, a) for a in A]
    print(*O)

if __name__ == "__main__":
    main()

# ########## PROGRAM: node_25:cc_python_25 ##########

from codebank import *

def main():
    import sys
    u, v = map(int, sys.stdin.readline().split())
    res = solve_xor_sum(u, v)
    if res is None:
        print(-1)
    else:
        print(len(res))
        if res:
            print(*res)

if __name__ == "__main__":
    main()

\end{lstlisting}
\end{figure}

\begin{figure}[H]
\begin{lstlisting}[style=codestyle, caption={Version 2, New Helpers}, label={helper:v2}]
# ==== NEW HELPER FUNCTIONS ====
def compute_complement(i):
    return i ^ ((1 << i.bit_length()) - 1)

def trie_add(trie, x, max_bit):
    trie[0][2] += 1
    now = 0
    for i in range(max_bit, -1, -1):
        bit = (x >> i) & 1
        if trie[now][bit] == 0:
            trie[now][bit] = len(trie)
            trie.append([0, 0, 0])
        now = trie[now][bit]
        trie[now][2] += 1

def trie_find_min_xor(trie, x, max_bit):
    now = 0
    ans = 0
    for i in range(max_bit, -1, -1):
        bit = (x >> i) & 1
        if trie[now][bit] and trie[trie[now][bit]][2] > 0:
            now = trie[now][bit]
        else:
            now = trie[now][bit ^ 1]
            ans |= (1 << i)
        trie[now][2] -= 1
    return ans

\end{lstlisting}
\end{figure}

\begin{figure}[H]
\begin{lstlisting}[style=codestyle, caption={Version 2, Refactored Programs}, label={code:v2}]
# ########## PROGRAM: node_16:cc_python_16 ##########

from codebank import *

def main():
    import sys
    input = sys.stdin.readline
    n = int(input())
    ans = [-1] * (n + 1)
    for i in range(n, -1, -1):
        if ans[i] == -1:
            z = compute_complement(i)
            ans[i] = z
            ans[z] = i
    m = sum(i ^ ans[i] for i in range(n + 1))
    print(m)
    print(*ans)

if __name__ == "__main__":
    main()

# ########## PROGRAM: node_19:cc_python_19 ##########

from codebank import *

def main():
    import sys
    input = sys.stdin.readline
    n = int(input())
    A = list(map(int, input().split()))
    P = list(map(int, input().split()))
    max_bit = max(max(A, default=0), max(P, default=0)).bit_length() - 1
    trie = [[0, 0, 0]]
    for x in P:
        trie_add(trie, x, max_bit)
    res = [trie_find_min_xor(trie, x, max_bit) for x in A]
    print(*res)

if __name__ == "__main__":
    main()

# ########## PROGRAM: node_25:cc_python_25 ##########

from codebank import *

def main():
    u, v = map(int, input().split())
    if u > v or ((v - u) & 1):
        print(-1)
    elif u == 0 and v == 0:
        print(0)
    elif u == v:
        print(1)
        print(u)
    else:
        w = (v - u) // 2
        if (w & u) == 0:
            d = u + w
            print(2)
            print(d, w)
        else:
            print(3)
            print(u, w, w)

if __name__ == "__main__":
    main()
\end{lstlisting}
\end{figure}

\subsection{Example 2}
In code snippets \ref{code:v1_2}, \ref{helper:v1_2}, \ref{code:v2_2}, \ref{helper:v2_2} is another example of 2 refactorings where V1 was better according to \NAME. We can observe that V2 creates helper functions that are overly specific to the problem. You can see that refactoring V2 introduces overly specialized functions like \texttt{dijkstra\_special()} or \texttt{compute\_min\_moves\_opposite\_parity()}. In comparison, refactoring V1 generates only general versions of these functions (e.g. \texttt{dijkstra()}).

\begin{figure}[H]
\begin{lstlisting}[style=codestyle, caption={Version 1, New Helpers}, label={helper:v1_2}]

# ==== NEW HELPER FUNCTIONS ====
def read_ints():
    return list(map(int, input().split()))

def build_adj_undirected(n, edges):
    adj = [[] for _ in range(n)]
    for u, v, w in edges:
        adj[u].append((v, w))
        adj[v].append((u, w))
    return adj

def dijkstra(adj, src):
    from heapq import heappush, heappop
    INF = 10**18
    n = len(adj)
    dist = [INF]*n
    parent = [-1]*n
    dist[src] = 0
    heap = [(0, src)]
    while heap:
        d, u = heappop(heap)
        if d > dist[u]:
            continue
        for v, w in adj[u]:
            nd = d + w
            if nd < dist[v]:
                dist[v] = nd
                parent[v] = u
                heappush(heap, (nd, v))
    return dist, parent

def reconstruct_path(parent, dest):
    path = []
    u = dest
    while u != -1:
        path.append(u+1)
        u = parent[u]
    return path[::-1]

def multi_source_bfs(neighbors, sources):
    from collections import deque
    n = len(neighbors)
    dist = [-1]*n
    dq = deque()
    for u in sources:
        if dist[u] == -1:
            dist[u] = 0
            dq.append(u)
    while dq:
        u = dq.popleft()
        for v in neighbors[u]:
            if dist[v] == -1:
                dist[v] = dist[u] + 1
                dq.append(v)
    return dist
\end{lstlisting}
\end{figure}

\begin{figure}[H]
\begin{lstlisting}[style=codestyle, caption={Version 1, Refactored Programs}, label={code:v1_2}]
# ########## PROGRAM: node_16:cc_python_16 ##########

from codebank import *

def main():
    import heapq
    n, m = read_ints()
    edges = [(u-1, v-1, w) for u, v, w in (read_ints() for _ in range(m))]
    adj = build_adj_undirected(n, edges)
    INF = 10**20
    dist = [INF]*n
    dist[0] = 0
    last_w = [0]*n
    heap = [(0, 0)]
    while heap:
        d, u = heapq.heappop(heap)
        if d > dist[u]:
            continue
        # record last edges
        for v, w in adj[u]:
            last_w[v] = w
        # expand two-edge moves
        for v, w1 in adj[u]:
            tw = last_w[v]
            for x, w2 in adj[v]:
                nd = d + (tw + w2)**2
                if nd < dist[x]:
                    dist[x] = nd
                    heapq.heappush(heap, (nd, x))
    out = []
    for x in dist:
        out.append(str(x if x < INF else -1))
    print(" ".join(out))

if __name__ == "__main__":
    main()

# ########## PROGRAM: node_17:cc_python_17 ##########

from codebank import *

def main():
    n, m = read_ints()
    edges = [(u-1, v-1, w) for u, v, w in (read_ints() for _ in range(m))]
    adj = build_adj_undirected(n, edges)
    dist, parent = dijkstra(adj, 0)
    if dist[n-1] >= 10**18:
        print(-1)
    else:
        path = reconstruct_path(parent, n-1)
        print(*path)

if __name__ == "__main__":
    main()

# ########## PROGRAM: node_19:cc_python_19 ##########

from codebank import *

def main():
    n = int(input())
    a = read_ints()
    # build reversed graph: for each move i->j, add edge j->i
    neighbors = [[] for _ in range(n)]
    for i, val in enumerate(a):
        for j in (i - val, i + val):
            if 0 <= j < n:
                neighbors[j].append(i)
    # BFS from all even and all odd positions separately
    even_sources = [i for i, val in enumerate(a) if val % 2 == 0]
    odd_sources  = [i for i, val in enumerate(a) if val % 2 == 1]
    dist_even = multi_source_bfs(neighbors, even_sources)
    dist_odd  = multi_source_bfs(neighbors, odd_sources)
    # for odd a[i], answer is dist to nearest even => dist_even; else dist_odd
    ans = [dist_even[i] if a[i] % 2 == 1 else dist_odd[i] for i in range(n)]
    print(*ans)

if __name__ == "__main__":
    main()
\end{lstlisting}
\end{figure}

\begin{figure}[H]
\begin{lstlisting}[style=codestyle, caption={Version 2, New Helpers}, label={helper:v2_2}]
#
# ==== NEW HELPER FUNCTIONS ====
def read_ints():
    return list(map(int, input().split()))

def build_undirected_weighted_graph(n, m):
    from collections import defaultdict
    adj = defaultdict(list)
    for _ in range(m):
        u, v, w = read_ints()
        u -= 1; v -= 1
        adj[u].append((v, w))
        adj[v].append((u, w))
    return adj

def dijkstra(adj, src, n):
    import heapq
    INF = 10**18
    dist = [INF]*n
    parent = [-1]*n
    visited = [False]*n
    dist[src] = 0
    heap = [(0, src)]
    while heap:
        d, u = heapq.heappop(heap)
        if visited[u]:
            continue
        visited[u] = True
        for v, w in adj.get(u, ()):
            nd = d + w
            if nd < dist[v]:
                dist[v] = nd
                parent[v] = u
                heapq.heappush(heap, (nd, v))
    return dist, parent

def reconstruct_path(parent, dest):
    path = []
    while dest != -1:
        path.append(dest+1)
        dest = parent[dest]
    return path[::-1]

def dijkstra_special(e, n, src):
    import heapq
    INF = 10**18
    d = [INF]*n
    d[src] = 0
    heap = [(0, src)]
    while heap:
        cd, v = heapq.heappop(heap)
        if cd > d[v]:
            continue
        td = {}
        for u, w in e.get(v, ()):
            td[u] = w
        for u, w1 in td.items():
            for x, w2 in e.get(u, ()):
                cost = cd + (w1 + w2)**2
                if cost < d[x]:
                    d[x] = cost
                    heapq.heappush(heap, (cost, x))
    return d

def compute_min_moves_opposite_parity(a):
    from collections import deque
    n = len(a)
    go = [[] for _ in range(n)]
    ans = [-1]*n
    q = deque()
    for i, val in enumerate(a):
        for j in (i - val, i + val):
            if 0 <= j < n:
                if (a[j] % 2) != (val % 2):
                    ans[i] = 1
                    q.append(i)
                    break
                else:
                    go[j].append(i)
    while q:
        u = q.popleft()
        for v in go[u]:
            if ans[v] == -1:
                ans[v] = ans[u] + 1
                q.append(v)
    return ans
\end{lstlisting}
\end{figure}

\begin{figure}[H]
\begin{lstlisting}[style=codestyle, caption={Version 2, Refactored Programs}, label={code:v2_2}]
#
# ########## PROGRAM: node_16:cc_python_16 ##########

from codebank import *

def main():
    n, m = read_ints()
    e = {}
    for _ in range(m):
        u, v, w = read_ints()
        u -= 1; v -= 1
        e.setdefault(u, []).append((v, w))
        e.setdefault(v, []).append((u, w))
    d = dijkstra_special(e, n, 0)
    print(" ".join(str(-1 if x >= 10**18 else int(x)) for x in d))

if __name__ == "__main__":
    main()

# ########## PROGRAM: node_17:cc_python_17 ##########

from codebank import *

def main():
    n, m = read_ints()
    adj = build_undirected_weighted_graph(n, m)
    dist, parent = dijkstra(adj, 0, n)
    if dist[n-1] >= 10**18:
        print(-1)
    else:
        path = reconstruct_path(parent, n-1)
        print(" ".join(map(str, path)))

if __name__ == "__main__":
    main()

# ########## PROGRAM: node_19:cc_python_19 ##########

from codebank import *

def main():
    n = int(input())
    a = read_ints()
    ans = compute_min_moves_opposite_parity(a)
    print(" ".join(map(str, ans)))

if __name__ == "__main__":
    main()

\end{lstlisting}
\end{figure}

\section{Obfuscation example}
We provide the full source code, both refactored and obfuscated, for the MDL and token comparison here.
\begin{lstlisting}[
    style=modernpython,
    caption={Refactored code from modeling\_llama.py from the refactored Transformers repository.}
]
from typing import Optional, Union

import torch
from torch import nn

from ...cache_utils import Cache, DynamicCache
from ...generation import GenerationMixin
from ...masking_utils import create_causal_mask
from ...modeling_layers import (
    GenericForQuestionAnswering,
    GenericForSequenceClassification,
    GenericForTokenClassification,
)
from ...modeling_outputs import (
    BaseModelOutputWithPast,
    CausalLMOutputWithPast,
)
from ...modeling_utils import PreTrainedModel
from ...processing_utils import Unpack
from ...utils import TransformersKwargs, auto_docstring, can_return_tuple, logging
from ...utils.generic import check_model_inputs
from .configuration_llama import LlamaConfig

from ..shared_library import (
    rotate_half,
    apply_rotary_pos_emb,
    repeat_kv,
    eager_attention_forward,
    RMSNorm,
    BaseMLP,
    BaseRotaryEmbedding,
    BaseAttention,
    BaseDecoderLayer,
)


logger = logging.get_logger(__name__)


class LlamaRMSNorm(RMSNorm):
    pass


class LlamaRotaryEmbedding(BaseRotaryEmbedding):
    pass


class LlamaMLP(BaseMLP):
    def __init__(self, config):
        super().__init__(config, mlp_bias=config.mlp_bias)


class LlamaAttention(BaseAttention):
    def __init__(self, config: LlamaConfig, layer_idx: int):
        super().__init__(
            config=config,
            layer_idx=layer_idx,
            attention_bias=config.attention_bias,
            sliding_window=None
        )


class LlamaDecoderLayer(BaseDecoderLayer):
    def __init__(self, config: LlamaConfig, layer_idx: int):
        super().__init__(
            config=config,
            layer_idx=layer_idx,
            norm_class=LlamaRMSNorm,
            mlp_class=LlamaMLP,
            attention_class=LlamaAttention
        )


@auto_docstring
class LlamaPreTrainedModel(PreTrainedModel):
    config: LlamaConfig
    base_model_prefix = "model"
    supports_gradient_checkpointing = True
    _no_split_modules = ["LlamaDecoderLayer"]
    _skip_keys_device_placement = ["past_key_values"]
    _supports_flash_attn = True
    _supports_sdpa = True
    _supports_flex_attn = True

    _can_compile_fullgraph = True
    _supports_attention_backend = True
    _can_record_outputs = {
        "hidden_states": LlamaDecoderLayer,
        "attentions": LlamaAttention,
    }


@auto_docstring
class LlamaModel(LlamaPreTrainedModel):
    def __init__(self, config: LlamaConfig):
        super().__init__(config)
        self.padding_idx = config.pad_token_id
        self.vocab_size = config.vocab_size

        self.embed_tokens = nn.Embedding(config.vocab_size, config.hidden_size, self.padding_idx)
        self.layers = nn.ModuleList(
            [LlamaDecoderLayer(config, layer_idx) for layer_idx in range(config.num_hidden_layers)]
        )
        self.norm = LlamaRMSNorm(config.hidden_size, eps=config.rms_norm_eps)
        self.rotary_emb = LlamaRotaryEmbedding(config=config)
        self.gradient_checkpointing = False

        self.post_init()

    @check_model_inputs
    @auto_docstring
    def forward(
        self,
        input_ids: Optional[torch.LongTensor] = None,
        attention_mask: Optional[torch.Tensor] = None,
        position_ids: Optional[torch.LongTensor] = None,
        past_key_values: Optional[Cache] = None,
        inputs_embeds: Optional[torch.FloatTensor] = None,
        cache_position: Optional[torch.LongTensor] = None,
        use_cache: Optional[bool] = None,
        **kwargs: Unpack[TransformersKwargs],
    ) -> BaseModelOutputWithPast:
        if (input_ids is None) ^ (inputs_embeds is not None):
            raise ValueError("You must specify exactly one of input_ids or inputs_embeds")

        if inputs_embeds is None:
            inputs_embeds: torch.Tensor = self.embed_tokens(input_ids)

        if use_cache and past_key_values is None:
            past_key_values = DynamicCache(config=self.config)

        if cache_position is None:
            past_seen_tokens = past_key_values.get_seq_length() if past_key_values is not None else 0
            cache_position: torch.Tensor = torch.arange(
                past_seen_tokens, past_seen_tokens + inputs_embeds.shape[1], device=inputs_embeds.device
            )

        if position_ids is None:
            position_ids = cache_position.unsqueeze(0)

        causal_mask = create_causal_mask(
            config=self.config,
            input_embeds=inputs_embeds,
            attention_mask=attention_mask,
            cache_position=cache_position,
            past_key_values=past_key_values,
            position_ids=position_ids,
        )

        hidden_states = inputs_embeds
        position_embeddings = self.rotary_emb(hidden_states, position_ids)

        for decoder_layer in self.layers[: self.config.num_hidden_layers]:
            hidden_states = decoder_layer(
                hidden_states,
                attention_mask=causal_mask,
                position_ids=position_ids,
                past_key_values=past_key_values,
                cache_position=cache_position,
                position_embeddings=position_embeddings,
                **kwargs,
            )

        hidden_states = self.norm(hidden_states)
        return BaseModelOutputWithPast(
            last_hidden_state=hidden_states,
            past_key_values=past_key_values,
        )


@auto_docstring
class LlamaForCausalLM(LlamaPreTrainedModel, GenerationMixin):
    _tied_weights_keys = ["lm_head.weight"]
    _tp_plan = {"lm_head": "colwise_rep"}
    _pp_plan = {"lm_head": (["hidden_states"], ["logits"])}

    def __init__(self, config):
        super().__init__(config)
        self.model = LlamaModel(config)
        self.vocab_size = config.vocab_size
        self.lm_head = nn.Linear(config.hidden_size, config.vocab_size, bias=False)

        self.post_init()

    def set_decoder(self, decoder):
        self.model = decoder

    def get_decoder(self):
        return self.model

    @can_return_tuple
    @auto_docstring
    def forward(
        self,
        input_ids: Optional[torch.LongTensor] = None,
        attention_mask: Optional[torch.Tensor] = None,
        position_ids: Optional[torch.LongTensor] = None,
        past_key_values: Optional[Cache] = None,
        inputs_embeds: Optional[torch.FloatTensor] = None,
        labels: Optional[torch.LongTensor] = None,
        use_cache: Optional[bool] = None,
        cache_position: Optional[torch.LongTensor] = None,
        logits_to_keep: Union[int, torch.Tensor] = 0,
        **kwargs: Unpack[TransformersKwargs],
    ) -> CausalLMOutputWithPast:
        r"""
        Example:

        ```python
        >>> from transformers import AutoTokenizer, LlamaForCausalLM

        >>> model = LlamaForCausalLM.from_pretrained("meta-llama/Llama-2-7b-hf")
        >>> tokenizer = AutoTokenizer.from_pretrained("meta-llama/Llama-2-7b-hf")

        >>> prompt = "Hey, are you conscious? Can you talk to me?"
        >>> inputs = tokenizer(prompt, return_tensors="pt")

        >>> # Generate
        >>> generate_ids = model.generate(inputs.input_ids, max_length=30)
        >>> tokenizer.batch_decode(generate_ids, skip_special_tokens=True, clean_up_tokenization_spaces=False)[0]
        "Hey, are you conscious? Can you talk to me?\nI'm not conscious, but I can talk to you."
        ```"""
        outputs: BaseModelOutputWithPast = self.model(
            input_ids=input_ids,
            attention_mask=attention_mask,
            position_ids=position_ids,
            past_key_values=past_key_values,
            inputs_embeds=inputs_embeds,
            use_cache=use_cache,
            cache_position=cache_position,
            **kwargs,
        )

        hidden_states = outputs.last_hidden_state
        slice_indices = slice(-logits_to_keep, None) if isinstance(logits_to_keep, int) else logits_to_keep
        logits = self.lm_head(hidden_states[:, slice_indices, :])

        loss = None
        if labels is not None:
            loss = self.loss_function(logits=logits, labels=labels, vocab_size=self.config.vocab_size, **kwargs)

        return CausalLMOutputWithPast(
            loss=loss,
            logits=logits,
            past_key_values=outputs.past_key_values,
            hidden_states=outputs.hidden_states,
            attentions=outputs.attentions,
        )


class LlamaForSequenceClassification(GenericForSequenceClassification, LlamaPreTrainedModel): ...


class LlamaForQuestionAnswering(GenericForQuestionAnswering, LlamaPreTrainedModel):
    base_model_prefix = "transformer"


class LlamaForTokenClassification(GenericForTokenClassification, LlamaPreTrainedModel): ...


__all__ = [
    "LlamaForCausalLM",
    "LlamaModel",
    "LlamaPreTrainedModel",
    "LlamaForSequenceClassification",
    "LlamaForQuestionAnswering",
    "LlamaForTokenClassification",
]
\end{lstlisting}

\begin{lstlisting}[
    style=modernpython,
    caption={Obfuscated code from modeling\_llama.py from the refactored Transformers repository.},
]
from typing import Optional as O,Union as U
import torch as T
from torch import nn as N
from ...cache_utils import Cache as C,DynamicCache as D
from ...generation import GenerationMixin as G
from ...masking_utils import create_causal_mask as m
from ...modeling_layers import GenericForQuestionAnswering as Q,GenericForSequenceClassification as S,GenericForTokenClassification as K
from ...modeling_outputs import BaseModelOutputWithPast as B,CausalLMOutputWithPast as L
from ...modeling_utils import PreTrainedModel as P
from ...processing_utils import Unpack as X
from ...utils import TransformersKwargs as W,auto_docstring as a,can_return_tuple as r,logging
from ...utils.generic import check_model_inputs as c
from .configuration_llama import LlamaConfig as F
from ..shared_library import rotate_half,apply_rotary_pos_emb,repeat_kv,eager_attention_forward,RMSNorm as R,BaseMLP as M,BaseRotaryEmbedding as E,BaseAttention as A,BaseDecoderLayer as Y
l=logging.get_logger(__name__)
class Z(R):pass
class I(E):pass
class J(M):
 def __init__(s,g):super().__init__(g,mlp_bias=g.mlp_bias)
class H(A):
 def __init__(s,g:F,i:int):super().__init__(config=g,layer_idx=i,attention_bias=g.attention_bias,sliding_window=None)
class V(Y):
 def __init__(s,g:F,i:int):super().__init__(config=g,layer_idx=i,norm_class=Z,mlp_class=J,attention_class=H)
@a
class o(P):
 config:F
 base_model_prefix="model"
 supports_gradient_checkpointing=True
 _no_split_modules=["LlamaDecoderLayer"]
 _skip_keys_device_placement=["past_key_values"]
 _supports_flash_attn=True
 _supports_sdpa=True
 _supports_flex_attn=True
 _can_compile_fullgraph=True
 _supports_attention_backend=True
 _can_record_outputs={"hidden_states":V,"attentions":H}
@a
class u(o):
 def __init__(s,g:F):
  super().__init__(g)
  s.padding_idx=g.pad_token_id
  s.vocab_size=g.vocab_size
  s.embed_tokens=N.Embedding(g.vocab_size,g.hidden_size,s.padding_idx)
  s.layers=N.ModuleList([V(g,i)for i in range(g.num_hidden_layers)])
  s.norm=Z(g.hidden_size,eps=g.rms_norm_eps)
  s.rotary_emb=I(config=g)
  s.gradient_checkpointing=False
  s.post_init()
 @c
 @a
 def forward(s,input_ids:O[T.LongTensor]=None,attention_mask:O[T.Tensor]=None,position_ids:O[T.LongTensor]=None,past_key_values:O[C]=None,inputs_embeds:O[T.FloatTensor]=None,cache_position:O[T.LongTensor]=None,use_cache:O[bool]=None,**k:X[W])->B:
  if(input_ids is None)^(inputs_embeds is not None):raise ValueError("You must specify exactly one of input_ids or inputs_embeds")
  if inputs_embeds is None:inputs_embeds:T.Tensor=s.embed_tokens(input_ids)
  if use_cache and past_key_values is None:past_key_values=D(config=s.config)
  if cache_position is None:
   p=past_key_values.get_seq_length()if past_key_values is not None else 0
   cache_position:T.Tensor=T.arange(p,p+inputs_embeds.shape[1],device=inputs_embeds.device)
  if position_ids is None:position_ids=cache_position.unsqueeze(0)
  f=m(config=s.config,input_embeds=inputs_embeds,attention_mask=attention_mask,cache_position=cache_position,past_key_values=past_key_values,position_ids=position_ids)
  h=inputs_embeds
  e=s.rotary_emb(h,position_ids)
  for d in s.layers[:s.config.num_hidden_layers]:h=d(h,attention_mask=f,position_ids=position_ids,past_key_values=past_key_values,cache_position=cache_position,position_embeddings=e,**k)
  h=s.norm(h)
  return B(last_hidden_state=h,past_key_values=past_key_values)
@a
class t(o,G):
 _tied_weights_keys=["lm_head.weight"]
 _tp_plan={"lm_head":"colwise_rep"}
 _pp_plan={"lm_head":(["hidden_states"],["logits"])}
 def __init__(s,g):
  super().__init__(g)
  s.model=u(g)
  s.vocab_size=g.vocab_size
  s.lm_head=N.Linear(g.hidden_size,g.vocab_size,bias=False)
  s.post_init()
 def set_decoder(s,d):s.model=d
 def get_decoder(s):return s.model
 @r
 @a
 def forward(s,input_ids:O[T.LongTensor]=None,attention_mask:O[T.Tensor]=None,position_ids:O[T.LongTensor]=None,past_key_values:O[C]=None,inputs_embeds:O[T.FloatTensor]=None,labels:O[T.LongTensor]=None,use_cache:O[bool]=None,cache_position:O[T.LongTensor]=None,logits_to_keep:U[int,T.Tensor]=0,**k:X[W])->L:
  o:B=s.model(input_ids=input_ids,attention_mask=attention_mask,position_ids=position_ids,past_key_values=past_key_values,inputs_embeds=inputs_embeds,use_cache=use_cache,cache_position=cache_position,**k)
  h=o.last_hidden_state
  i=slice(-logits_to_keep,None)if isinstance(logits_to_keep,int)else logits_to_keep
  g=s.lm_head(h[:,i,:])
  n=None
  if labels is not None:n=s.loss_function(logits=g,labels=labels,vocab_size=s.config.vocab_size,**k)
  return L(loss=n,logits=g,past_key_values=o.past_key_values,hidden_states=o.hidden_states,attentions=o.attentions)
class b(S,o):...
class x(Q,o):base_model_prefix="transformer"
class y(K,o):...
__all__=["t","u","o","b","x","y"]
\end{lstlisting}

\end{document}